%% file: PASSAGE_COSMOS_FINAL.tex
\documentclass[twocolumn]{aastex701}
\usepackage[utf8]{inputenc}
\usepackage[figure,figure*]{hypcap}
\usepackage{multirow}
\usepackage{natbib}
\usepackage{xcolor}
\usepackage{soul}
\usepackage{CJK}
\usepackage{enumitem}
\usepackage{xspace}
\usepackage{amsmath}
\usepackage{comment}

\setstcolor{red}
\hypersetup{linkcolor=red,citecolor=cyan,filecolor=magenta,urlcolor=cyan}

\newcommand{\Lya}{\textrm{Ly}\ensuremath{\alpha}\xspace}
\newcommand{\Ha}{\textrm{H}\ensuremath{\alpha}\xspace}
\newcommand{\Hb}{\textrm{H}\ensuremath{\beta}\xspace}
\newcommand{\Paa}{\textrm{Pa}\ensuremath{\alpha}\xspace}
\newcommand{\Pab}{\textrm{Pa}\ensuremath{\beta}\xspace}
\newcommand{\Pag}{\textrm{Pa}\ensuremath{\gamma}\xspace}
\newcommand{\Hg}{\textrm{H}\ensuremath{\gamma}\xspace}
\newcommand{\Hd}{\textrm{H}\ensuremath{\delta}\xspace}
\newcommand{\Heps}{\textrm{H}\ensuremath{\epsilon}\xspace}

\newcommand{\OI}{[\textrm{O}~\textsc{i}]\xspace}
\newcommand{\OII}{[\textrm{O}~\textsc{ii}]\xspace}
\newcommand{\OIII}{[\textrm{O}~\textsc{iii}]\xspace}
\newcommand{\CIII}{\textrm{C}~\textsc{iii}]\xspace}
\newcommand{\NII}{[\textrm{N}~\textsc{ii}]\xspace}
\newcommand{\Nv}{\textrm{N}~\textsc{v}\xspace}
\newcommand{\SII}{[\textrm{S}~\textsc{ii}]\xspace}
\newcommand{\SIII}{[\textrm{S}~\textsc{iii}]\xspace}
\newcommand{\MgII}{\textrm{Mg}~\textsc{ii}\xspace}
\newcommand{\SiIII}{\textrm{Si}~\textsc{iii}]\xspace}
\newcommand{\NeIII}{[\textrm{Ne}~\textsc{iii}]\xspace}
\newcommand{\HeII}{\textrm{He}~\textsc{ii}\xspace}
\newcommand{\HeI}{\textrm{He}~\textsc{i}\xspace}
\newcommand{\CIV}{\textrm{C}~\textsc{iv}\xspace}

\newcommand{\grizli}{{\sc Grizli}}
\newcommand{\sep}{{\sc SEP}}

\newcommand\tnmark[1]{\tablenotemark{\scriptsize {#1}}}
\newcommand\tntext[1]{\tablenotetext{{#1}}}

\begin{document}

\title{The NIRISS PASSAGE Spectroscopic Redshift Catalog in COSMOS}

\author[0009-0002-9932-4461]{Mason S. Huberty}
\affiliation{Minnesota Institute for Astrophysics, University of Minnesota, Twin Cities, 116 Church St SE, Minneapolis, MN 55455, USA}
\email{huber458@umn.edu}  

\author[0000-0001-5294-8002]{Kalina~V.~Nedkova}
\affiliation{IPAC, California Institute of Technology, 1200 E.~California Blvd, Pasadena, CA 91125, USA}
\affiliation{Department of Physics and Astronomy, Johns Hopkins University, 3400 North Charles Street, Baltimore, MD 21218, USA}
\affiliation{Space Telescope Science Institute, 3700 San Martin Drive, Baltimore, MD 21218, USA}
\email{knedkova@stsci.edu}

\author[0000-0002-0364-1159]{Zahra Sattari}
\affiliation{IPAC, California Institute of Technology, 1200 E.~California Blvd, Pasadena, CA 91125, USA}
\email{zsattari@ipac.caltech.edu}

\author[0000-0001-7166-6035]{Vihang Mehta}
\affiliation{IPAC, California Institute of Technology, 1200 E.~California Blvd, Pasadena, CA 91125, USA}
\email{vmehta@ipac.caltech.edu}

\author[0000-0002-9136-8876]{Claudia Scarlata}
\affiliation{Minnesota Institute for Astrophysics, University of Minnesota, Twin Cities, 116 Church St SE, Minneapolis, MN 55455, USA}
\email{mscarlat@umn.edu} 

\author[0000-0002-9946-4731]{Marc Rafelski}
\affiliation{Space Telescope Science Institute, 3700 San Martin Drive, Baltimore, MD 21218, USA}
\affiliation{Department of Physics and Astronomy, Johns Hopkins University, 3400 North Charles Street, Baltimore, MD 21218, USA}
\email{mrafelski@stsci.edu}

\author[0000-0001-8587-218X]{Matthew J. Hayes}
\affiliation{Stockholm University, Department of Astronomy, AlbaNova University Center, SE-106 91 Stockholm, Sweden}
\email{matthew@astro.su.se} 

\author[0000-0003-3108-0624]{Peter J. Watson}
\affiliation{INAF, Osservatorio Astronomico di Padova, Vicolo dell'Osservatorio 5, 35122 Padova, Italy}
\email{peter.watson@inaf.it}

\author[0000-0003-4804-7142]{Ayan Acharyya}
\affiliation{INAF, Osservatorio Astronomico di Padova, Vicolo dell'Osservatorio 5, 35122 Padova, Italy}
\email{ayan.acharyya@inaf.it}

\author[0009-0000-9478-1933]{Jacob Levine}
\affiliation{University of California, Los Angeles, Department of Physics and Astronomy, 430 Portola Plaza, Los Angeles, CA 90095, USA}
\email{fhasan@stsci.edu}

\author[0000-0003-0980-1499]{Benedetta Vulcani}
\affiliation{INAF, Osservatorio Astronomico di Padova, Vicolo dell'Osservatorio 5, 35122 Padova, Italy}
\email{benedetta.vulcani@inaf.it}

\author[0000-0003-1767-6421]{Alexandra Le Reste}
\affiliation{Minnesota Institute for Astrophysics, University of Minnesota, Twin Cities, 116 Church St SE, Minneapolis, MN 55455, USA}
\email{alereste@umn.edu}

\author[0000-0002-0072-0281]{Farhanul Hasan}
\affiliation{Space Telescope Science Institute, 3700 San Martin Drive, Baltimore, MD 21218, USA}
\email{fhasan@stsci.edu}

\author[0000-0001-6482-3020]{James Colbert}
\affiliation{IPAC, California Institute of Technology, 1200 E.~California Blvd, Pasadena, CA 91125, USA}
\email{colbert@ipac.caltech.edu}

\author[0000-0001-9391-305X]{Michele Trenti}
\affiliation{School of Physics, University of Melbourne, Parkville 3010, VIC, Australia}
\affiliation{ARC Centre of Excellence for All Sky Astrophysics in 3 Dimensions (ASTRO 3D), Australia}
\email{michele.trenti@unimelb.edu.au }

\author[0000-0002-9373-3865]{Xin Wang}
\affiliation{School of Astronomy and Space Science, University of Chinese Academy of Sciences (UCAS), Beijing 100049, China}
\affiliation{National Astronomical Observatories, Chinese Academy of Sciences, Beijing 100101, China}
\affiliation{Institute for Frontiers in Astronomy and Astrophysics, Beijing Normal University, Beijing 102206, China}
\email{xwang@ucas.ac.cn}

\author[0000-0002-1025-7569]{Axel Runnholm}
\affiliation{Stockholm University, Department of Astronomy, AlbaNova University Center, SE-106 91 Stockholm, Sweden}
\email{} 

\author[0000-0001-6919-1237]{Matthew A. Malkan}
\affiliation{University of California, Los Angeles, Department of Physics and Astronomy, 430 Portola Plaza, Los Angeles, CA 90095, USA}
\email{malkan@astro.ucla.edu}

\author[0000-0002-8651-9879]{Andrew J.\ Bunker}
\affiliation{University of Oxford, Department of Physics, Keble Road, Oxford OX1 3RH, UK}
\email{andy.bunker@physics.ox.ac.uk}

\author[0000-0002-8630-6435]{Anahita Alavi}
\affiliation{IPAC, California Institute of Technology, 1200 E. California Blvd, Pasadena, CA 91125, USA}
\email{anahita@ipac.caltech.edu}

\author[0000-0002-7570-0824]{Hakim Atek}
\affiliation{CNRS, Institut d'Astrophysique de Paris, 98 bis Boulevard Arago, 75014 Paris, France}
\email{atek@iap.fr}

\author[0000-0003-4569-2285]{Andrew J. Battisti}
\affiliation{International Centre for Radio Astronomy Research, University of Western Australia, 7 Fairway, Crawley, WA 6009, Australia}
\affiliation{Research School of Astronomy and Astrophysics, Australian National University, Cotter Road, Weston Creek, ACT 2611, Australia}
\email{ajbattisti@gmail.com}

\author[0000-0002-7928-416X]{Y. Sophia Dai}
\affiliation{Chinese Academy of Sciences South America Center for Astronomy (CASSACA), National Astronomical Observatories(NAOC),
20A Datun Road, Beijing 100012, China}
\email{ydai@nao.cas.cn}

\author[0000-0001-6505-0293]{Keunho Kim}
\affiliation{IPAC, California Institute of Technology, 1200 E.~California Blvd, Pasadena, CA 91125, USA}
\email{keunho11@ipac.caltech.edu}

\author[0000-0002-6586-4446]{Alaina Henry}
\affiliation{Space Telescope Science Institute, 3700 San Martin Drive, Baltimore, MD 21218, USA}
\email{ahenry@stsci.edu} 

\author[0000-0001-7016-5220]{Michael J. Rutkowski}
\affiliation{Minnesota State University, Mankato, Department of Physics and Astronomy, 141 Trafton Science Center N, Mankato, MN 56001, USA}
\email{}

\author[0000-0003-3596-8794]{Hollis Akins}
\affiliation{Department of Physics, University of California, Santa Barbara, Santa Barbara, CA 93106, USA}
\email{hollis.akins@gmail.com}

\author[0000-0002-0930-6466]{Caitlin M. Casey}
\affiliation{Department of Physics, University of California, Santa Barbara, Santa Barbara, CA 93106, USA}
\affiliation{The University of Texas at Austin, 2515 Speedway Blvd Stop C1400, Austin, TX 78712, USA}
\affiliation{Cosmic Dawn Center (DAWN), Denmark}
\email{cmcasey.astro@gmail.com}

\author[0000-0002-3560-8599]{Maximilien Franco}
\affiliation{Université Paris-Saclay, Université Paris Cité, CEA, CNRS, AIM, 91191 Gif-sur-Yvette, France}
\email{}

\author[0000-0003-0129-2079]{Santosh Harish}
\affiliation{Laboratory for Multiwavelength Astrophysics, School of Physics and Astronomy, Rochester Institute of Technology, 84 Lomb Memorial Drive, Rochester, NY 14623, USA}
\email{harish.santosh@gmail.com}

\author[0000-0001-9187-3605]{Jeyhan S. Kartaltepe}
\affiliation{Laboratory for Multiwavelength Astrophysics, School of Physics and Astronomy, Rochester Institute of Technology, 84 Lomb Memorial Drive, Rochester, NY 14623, USA}
\email{jsksps@rit.edu}

\author[0000-0002-6610-2048]{Anton Koekemoer}
\affiliation{Space Telescope Science Institute, 3700 San Martin Drive, Baltimore, MD 21218, USA}
\email{koekemoer@stsci.edu}

\author[0000-0001-9773-7479]{Daizhong Liu}
\affiliation{Purple Mountain Observatory, Chinese Academy of Sciences, 10 Yuanhua Road, Nanjing 210023, China}
\email{}

\author[0000-0002-9489-7765]{Henry McCracken}
\affiliation{Institut d’Astrophysique de Paris, UMR 7095, CNRS, and Sorbonne Université, 98 bis boulevard Arago, F-75014 Paris, France}
\email{}

\author[0000-0002-4485-8549]{Jason Rhodes}
\affiliation{Jet Propulsion Laboratory, California Institute of Technology, 4800 Oak Grove Drive, Pasadena, CA 91001, USA}
\email{jason.d.rhodes@jpl.nasa.gov}
\author[0000-0002-4271-0364]{Brant Robertson}
\affiliation{Department of Astronomy and Astrophysics, University of California, Santa Cruz, 1156 High Street, Santa Cruz, CA 95064, USA}
\email{brant@ucsc.edu}

\author[0000-0002-7087-0701]{Marko Shuntov}
\affiliation{Cosmic Dawn Center (DAWN), Denmark} 
\affiliation{Niels Bohr Institute, University of Copenhagen, Jagtvej 128, DK-2200, Copenhagen, Denmark}
\email{marko.shuntov@nbi.ku.dk}

\correspondingauthor{Mason S. Huberty}
\email{huber458@umn.edu}

\begin{abstract}
We present the Parallel Application of Slitless Spectroscopy to Analyze Galaxy Evolution (PASSAGE) spectroscopic redshift catalog in the COSMOS field. PASSAGE is a JWST Cycle 1 Near Infrared Imager and Slitless Spectrograph (NIRISS) wide-field slitless spectroscopy (WFSS) pure-parallel survey, obtaining near-infrared spectra of thousands of extragalactic sources. 15 out of 63 PASSAGE fields fall within the Hubble Space Telescope (HST) COSMOS footprint, of which 11 overlap with COSMOS-Web, a JWST treasury survey providing additional space-based photometry. 
We present our custom line-finding algorithm and visual inspection effort used to identify emission lines and derive the spectroscopic redshifts for line-emitting sources in PASSAGE. The line-finding algorithm identifies between $\sim200$ and $950$ line-emitting candidates per field, of which typically $47\%$  were identified as true emission lines post visual inspection. 
We identify 2183 emission line sources at $0.08\lesssim z\lesssim 4.7$, 1896 of which have available COSMOS photometric redshifts. We find excellent redshift agreement between the COSMOS photometric redshifts and the PASSAGE spectroscopic redshifts for strong (S/N$>5$), multi-line emitting sources.
This agreement weakens for PASSAGE single-line emitters with ambiguous identities. These single-line emitters are likely mis-identified around $18\%$ of the time based on comparisons to photometric redshifts. We derive stellar masses using PASSAGE photometry and spectroscopic redshifts, in broad agreement with existing COSMOS-Web stellar masses, but with some discrepancy driven by redshift disagreements.
We publicly release this spectroscopic redshift catalog, which will enable community-led science in prime extragalactic fields and serve as a crucial dataset for validating Euclid and Roman spectroscopy.
\end{abstract}

\keywords{catalogs --- galaxies: redshifts --- galaxies: fundamental parameters --- methods: observational --- techniques: spectroscopic}

\section{Introduction}
\label{sec:intro}
The James Webb Space Telescope (JWST) has opened up a new parameter space in the study of galaxy evolution. 
JWST's wide-field slitless spectroscopy (WFSS) observations provide unbiased observations of thousands of sources simultaneously by obtaining spectra for all objects in the field of view without target pre-selection.
The unparalleled sensitivity of JWST in the near-infrared (NIR) permits the detection of faint emission lines for both known and new extragalactic sources. 
Galaxies in the mid- to low-mass regime ($\rm log[M_*/M_\odot] \lesssim 10$) are of particular interest, as they represent a larger share of the galaxy population than their massive counterparts -- yet prior to JWST, they could not be reliably characterized at moderate to high redshifts.
Optical emission lines such as the Balmer lines, \OIII\ 4960, 5008~{\AA}, and \OII 3727, 3730~{\AA} from sources at $1\lesssim z\lesssim 4$ fall in the NIR regime, alongside \SIII\ 9069, 9532~{\AA} and Paschen lines at the lowest redshifts ($z<1$). These lines can be used to very precisely constrain the redshift of their hosts and enable measurements of their physical properties.
JWST WFSS surveys have already proved fruitful, providing insights into the baryon cycle, reionization and the early universe, dwarf galaxy physical properties, large scale structure, and AGN, e.g., EIGER \citep{kashino2023}, FRESCO \citep{oesch2023}, CEERS \citep{finkelstein2025}, CANUCS \citep{sarrouh2025}, and GLASS \citep{treu2022,watson2025}. 

The Near Infrared Imager and Slitless Spectrograph (NIRISS) provides WFSS in the NIR wavelength range of $\sim1$ to $\sim2.2\mu \rm m$ \citep{willott2022}. 
While the modest resolution of NIRISS ($\rm R\sim150$) may limit the number of spectral features that can be identified, it is well-suited for observations of large populations of sources, the identification of strong emission lines, spatially resolved studies \citep[e.g.,][]{Estrada-Carpenter2025ApJ, acharyya2025}, and high-redshift galaxy detection \citep[e.g.,][]{runnholm2025}. 
 
The Parallel Application of Slitless Spectroscopy
to Analyze Galaxy Evolution (PASSAGE) survey is a Cycle 1 JWST GO pure-parallel survey utilizing the NIRISS WFSS mode \citep{Malkan.2025}. Owing to its pure-parallel design, the survey has pseudo-random pointings determined by independent JWST primary observations, with NIRISS data collected simultaneously. As a result, the effects of cosmic variance are substantially mitigated, since each PASSAGE field is independent of the others.

That said, a significant number of PASSAGE fields fall within the Cosmic Evolution Survey \citep[COSMOS,][]{scoville2007} field, as a result of broad observational attention devoted to this region of the sky. COSMOS is a $\sim$2 square degree field with extensive multi-wavelength data \citep{scoville2007}, acting as a crucial training field for JWST surveys. Observations of the COSMOS field began with Hubble Space Telescope (HST) Advanced Camera Survey (ACS) F814W \citep{koekemoer2007} and continued with Spitzer Space Telescope mid-infrared imaging \citep{sanders2007}, VISTA NIR coverage \citep{mccracken2012}, Subaru/Supremce Cam imaging \citep{capak2007,taniguchi2015}, the Canada-France-Hawaii Telescope program CLAUDS \citep{sawicki2019}, and the Hyper-Supreme Cam (HSC) Subaru Strategic Program (SSP) \citep{aihara2019}, leading to the COSMOS2020 catalog \citep{weaver2022}. 
With the launch of JWST, COSMOS2020 was expanded even further to the COSMOS-Web/COSMOS2025 catalog with the addition of JWST NIRCAM (F115W, F150W, F277W, and F444W) and MIRI (F770W) filters \citep{shuntov2025}.
This extensive photometric dataset enables well-constrained SED modeling, yielding robust galaxy properties such as stellar masses, which cannot be reliably derived from the wavelength-limited NIRISS spectroscopic data alone. 
The COSMOS multi-band photometry also enables reliable photometric redshift estimates for many sources in the field.
However, these redshifts often carry large uncertainties, making spectroscopic redshifts essential for accurately constraining galaxy distances.
Sources in the overlap between PASSAGE and COSMOS therefore benefit from robust spectroscopic redshifts and extensive photometric coverage, together yielding a large, well-constrained sample for studies of galaxy evolution. Although COSMOS already hosts one of the most comprehensive spectroscopic catalogs  \citep[see][]{khostovan2025}, the majority of previous surveys relied on pre-selection and are consequently biased toward more massive galaxies. PASSAGE is complementary by providing spectroscopic redshifts for lower-mass systems.

In this paper, we present the %PASSAGE in COSMOS 
spectroscopic redshift and stellar mass catalog for the 15 PASSAGE fields that overlap with COSMOS HST ACS F814W imaging, including the extensive emission line-finding effort and visualization tools that are required for such a study.
This paper is structured as follows: in section~\ref{sec:sample}, we present the PASSAGE sample and data reduction procedures. In section~\ref{sec:emissionlineidentification}, we present the emission line-finding procedure. 
In section~\ref{sec:finalresults}, we present the PASSAGE in COSMOS catalog. 
We conduct redshift comparisons of the PASSAGE sample to ancillary data in section~\ref{sec:discussion}.
We discuss the statistics of the catalog in  section~\ref{sec:redstats}. The stellar masses of the sample are discussed in section~\ref{sec:stellarmasses}. Our conclusions are outlined in section~\ref{sec:conclusion}.

In this paper, the standard concordance cosmology model ($\Omega_{\rm m}=0.3$, $\Omega_{\rm \Lambda}=0.7$, $H_0=70~{\rm km/s/Mpc}$) and AB magnitude system \citep{oke1983} are used.
Forbidden lines are indicated as follows, if presented without wavelength values:
$\OII\lambda\lambda3727,3730=\OII$, $\NeIII~\lambda3869=\NeIII$, 
$\OIII\lambda5008=\OIII$, $\NII~\lambda6585=\NII$, $\SII~\lambda\lambda6718,6732=\SII$.

\section{The PASSAGE Survey}\label{sec:sample}

PASSAGE is a Cycle 1 JWST GO pure-parallel survey, that obtained NIRISS imaging and slitless spectroscopy of 63 fields \citep[totaling $\sim 305$ square arcminutes,][]{Malkan.2025}.

PASSAGE is characterized by several key features: it is an unbiased spectroscopic survey without photometric pre-selection; it samples dozens of independent sightlines, mitigating the effects of cosmic variance; it provides continuous NIR coverage over $\sim1$–$2.2,\mu\mathrm{m}$, capturing rest-frame optical emission lines at cosmic noon; and it combines high spatial resolution (0\farcs065 per pixel) with up to two orthogonal grism orientations to reduce spectral overlap and contamination. In total, PASSAGE has delivered over 10,000 NIR grism spectra across these fields, enabling statistically robust studies of galaxy evolution across a wide range of environments and redshifts.

The PASSAGE observing strategy is thoroughly discussed in \citet{Malkan.2025}, but to briefly summarize: 
in each field, NIRISS spectroscopy is taken with up to 3 filters (F115W, F150W, and F200W) and up to 2 orthogonal grism orientations (GR150R and GR150C). Direct imaging is also obtained for each field in up to 3 filters (F115W, F150W, F200W). 
PASSAGE prioritizes maximizing spectroscopic coverage such that some fields have multi-grism filter coverage but direct imaging in only 1 or 2 bands.

Exposure times for the survey fields vary from $\sim1$ to $\sim20$ hours, allowing for emission line flux depths reaching between $10^{-17}$ and $10^{-18}$ cgs.

\begin{figure*}\label{fig:sensitivity}
\includegraphics[width=1\linewidth]{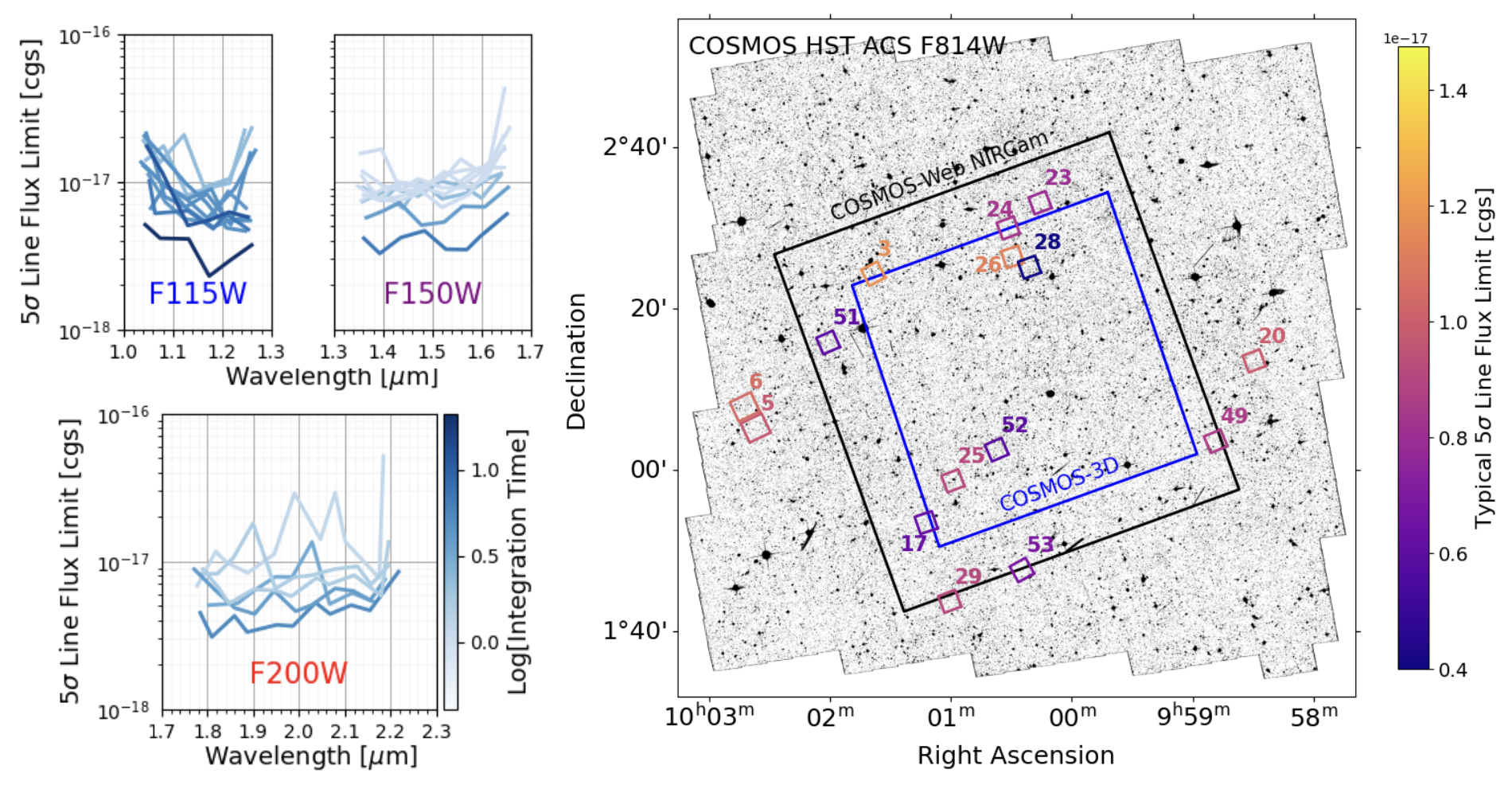}
\caption{Left panels: typical PASSAGE 5$\sigma$ detection limits for emission lines in PASSAGE (including both point-sources and extended sources), in erg/sec/cm$^{2}$ for the F115W (upper-left), F150W (upper-right), and F200W (bottom). The color of the lines corresponding to their integration times, with a darker color indicating a longer integration time.
Right panel: locations of the 15 PASSAGE fields analyzed in this work, relative to the COSMOS HST (the base of the COSMOS2020 catalog) \citep{koekemoer2007,massey2010}, COSMOS-Web \citep{franco2025}, and COSMOS-3D \citep{kakiichi2024} footprints. 
The color of each field corresponds with its typical 5$\sigma$ detection limit.} 
\end{figure*}

\subsection{PASSAGE Fields in COSMOS}
15 out of the 63 PASSAGE fields lie within the COSMOS2020 HST/ACS footprint, as shown in the right panel of Figure~\ref{fig:sensitivity}. 
The left panels of Figure~\ref{fig:sensitivity} show the typical emission-line flux sensitivity as a function of wavelength for each of these 15 fields (not all fields have all 3 filters), including both point and extended sources (emission lines from extended source tend to have weaker significance than point sources). The color of each line encodes the integration time of the corresponding exposure, with darker shades indicating longer exposures. % CS: I DON'T THINK WE NEED THIS, AS THE FIGURE DOESN'T SHOW THE PARS ANYMORE: Par028 (Par $=$ parallel field), the field with the longest exposure time in all 3 filters, is naturally the most sensitive field. 
For the 6 fields that have observations in both orientations, the total exposure time is split approximately evenly between the two orientations. 

These 15 fields and their observational properties are summarized in Table~\ref{tab:obs_table}. Eight fields lie entirely within the COSMOS-Web footprint, three partially overlap COSMOS-Web (with the remaining area covered by the broader COSMOS2020 footprint), and four additional fields fall outside COSMOS-Web but remain within COSMOS2020.

Two fields have spectroscopy in all three filters (F115W, F150W, and F200W), eight fields were observed in two filters (F115W and F150W), and the remaining five were observed in a single filter, typically F200W (with the exception of Par051, which has only F115W coverage). Six fields include spectroscopy in at least one grism filter obtained at two orthogonal orientations using both GR150R and GR150C. Among these, Par028 stands out as one of the most complete PASSAGE fields, lying within COSMOS-Web and featuring the deepest observations in all three filters, each obtained at both orientations \citep[see e.g.,][]{acharyya2025}. In addition, several PASSAGE fields overlap with the JWST Cycle~3 COSMOS-3D survey \citep{kakiichi2024}, which provides complementary NIRCam F444W WFSS data and imaging in F115W, F200W, and F356W.

\subsection{Spectral Extraction}\label{sec:passagedatareduction}
Slitless spectral extraction and contamination modeling are performed using the Grism Redshift \&\ Line analysis software \citep[\grizli\footnote{\url{https://grizli.readthedocs.io}, version 1.12.11}; ][]{Grizli:2021}. Redshift measurements are carried out separately as described below.
A detection image for each field is made by taking the weighted mean of all direct images for all filters that are available for that field. 
All sources in the detection image are identified using the Python Library for Source Extraction and Photometry \citep[\sep\footnote{\url{https://sep.readthedocs.io}}][]{barbary2016,bertin1996}. 
\sep\ typically identifies between $\sim3000-4500$ sources in each field and is provided in Table~\ref{tab:obs_table}. \grizli\ also measures photometric fluxes for each filter with a direct image. 

In order to model the overlapping spectra, \grizli\ takes the segmentation map of the \sep-identified sources and models their associated spectra in each individual grism exposure,
providing 2D and 1D spectra of each source for all available filters and orientations in each field. 
Combined 1D and 2D spectra are also generated for filters with multiple orientations.

Unfortunately, the spectral extraction by \grizli\ contains some noise and diffraction spikes that are incorrectly identified as extragalactic sources. 
Additionally, when \grizli\ processes the spectra of two overlapping traces, the automatic removal of the contamination is not perfect, occasionally resulting in an over-subtraction. 
These errors can result in the identification of fictitious extragalactic sources and spurious emission lines. 
Therefore, human intervention is essential to ensure an accurate sample of emission line sources.

\input{obs_table.tbl}

\section{Emission Line Identification}\label{sec:emissionlineidentification}
Emission lines are identified in the \grizli-extracted spectra using a line-finding pipeline that performs automated detection of emission-line candidates, followed by visual inspection of all candidates. The code is made publicly available on GitHub\footnote{\url{https://github.com/jwstwfss/line-finding}} \citep{nedkova2026}, and the automated and visual components of the pipeline are described in Sections~\ref{sec:AutomatedLineDetection} and \ref{sec:VisualLineDetection}, respectively. This pipeline has also been applied in other JWST WFSS surveys \citep{Battisti:2024,poppies2024}.
%comment

\subsection{Automated Line Detection} \label{sec:AutomatedLineDetection}
As part of this paper, we release the line-finding software used to identify emission lines and measure spectroscopic redshifts. Emission line candidates in the reduced 1D spectra are identified using a continuous wavelet transform (CWT) that selects appropriately-shaped peaks. This approach is described in full detail in \cite{Battisti:2024} where it was used on HST/WFC3 WFSS data, and is based on the \texttt{find\_peaks\_cwt} routine from \cite{Du:2006}. 

For completeness, we briefly summarize the algorithm here: 
similar to a Fourier transform, the wavelet transform breaks the \grizli\ extracted 1D signal into base constituents. However, rather than decomposing the signal into sinusoidal components like a Fourier transform, the base components are a scaled and shifted version of a `mother' Ricker wavelet--a function that is proportional to the second derivative of the Gaussian function. The CWT transform compares the wavelet with the PASSAGE 1D spectrum, shifting the wavelet along the spectrum's wavelength range and scaling (i.e. stretching/compressing) the wavelet. This creates a matrix of CWT coefficients that represent the correlation between the 1D spectrum and the wavelets at all wavelengths and across all scales. Emission-line candidates correlate with the wavelets on many scales, and therefore can be identified as strong `ridges' in the CWT coefficient matrix.

Thus, emission line candidates are identified based on both their amplitude and shape, reducing the amount of spurious emission lines that need to be visually inspected.

\begin{figure*}
    \centering
    \includegraphics[width=1\linewidth]{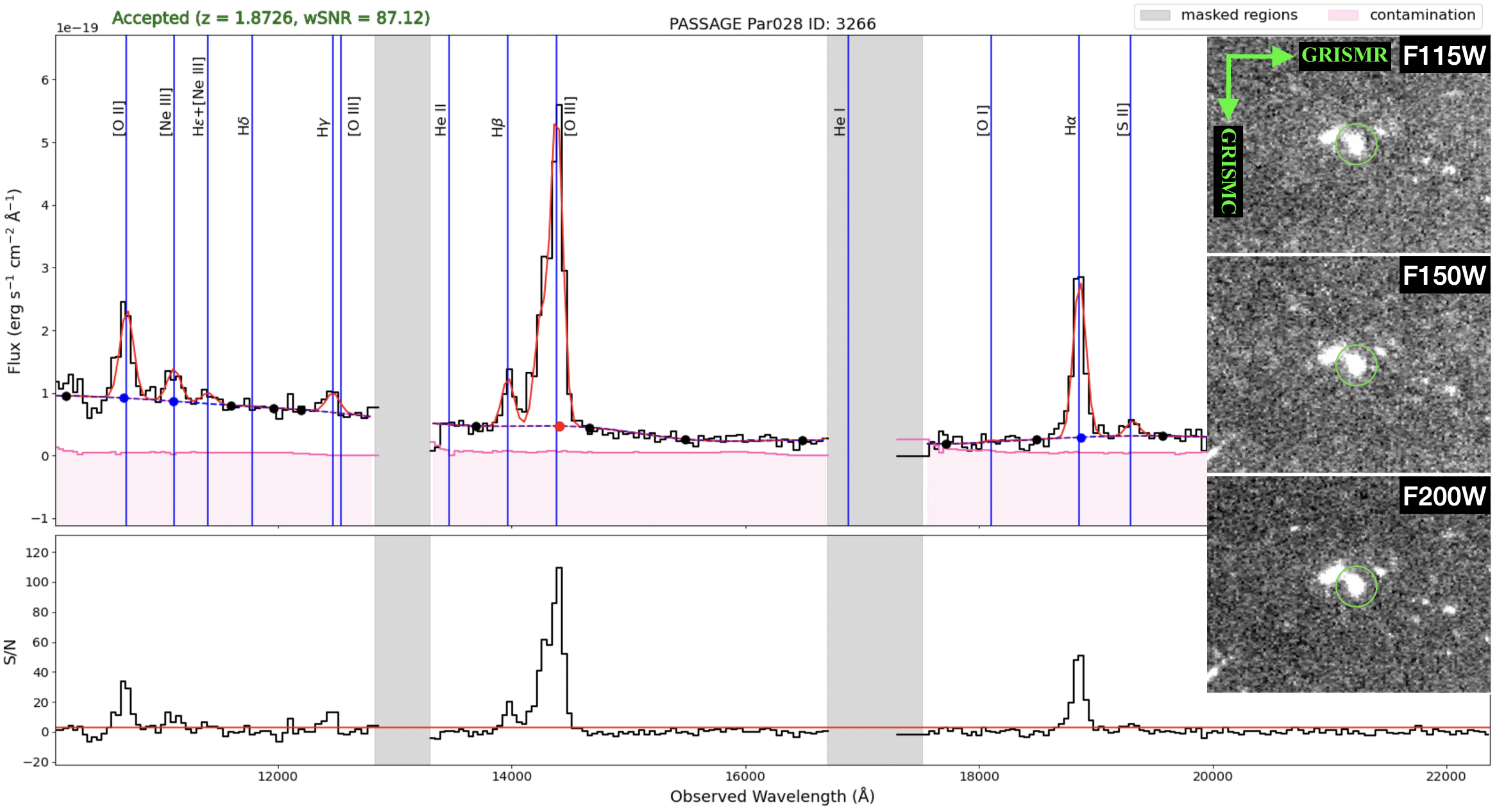}
    \includegraphics[width=1\linewidth]{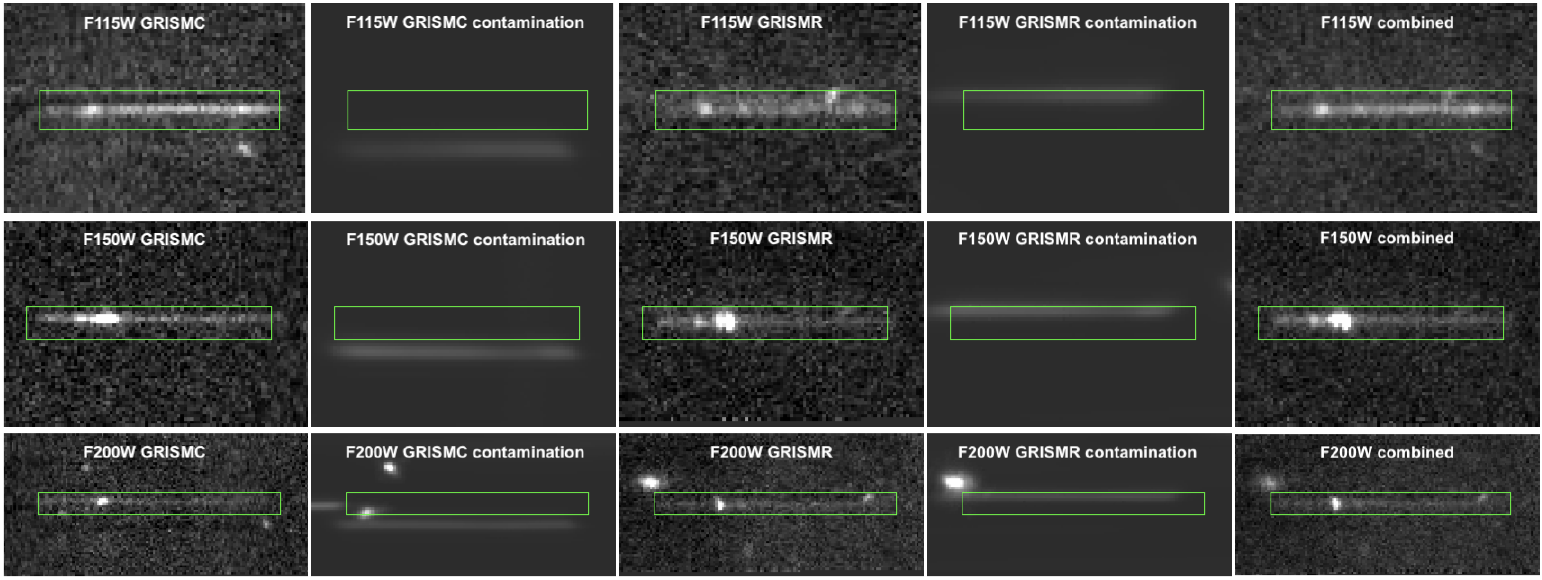}
    \caption{Visual interface of the emission line-finding algorithm for an example galaxy at $z=1.87$ in Par028 seen in all three filters in both grism orientations. The top panel shows the combined 1D spectrum for the galaxy (black), the contamination model (pink), Gaussian fits of emission lines (red) corresponding to the observed wavelength of the lines marked as vertical blue lines. Gaps between the grism filters are shown in gray. The SNR of the spectrum at each wavelength is shown in the panel below the 1D spectrum. 
    Cutouts of the direct image in F115W, F150W, and F200W are shown in the top-right, with the target source circle in green. The green arrows represent the dispersion direction for the two GRISM orientations.
    The three bottom rows show the 2D spectra for this object: the top row shows the F115W spectra, the middle row shows the F150W spectra, and the bottom row shows the F200W spectra. The first column shows the filters in the GRISMC orientation, the third column shows the GRISMR orientation, and the final column shows the orientation-combined 2D spectrum. The second and fourth columns show the GRISMC and GRISMR contamination models, respectively, marking spectral traces from neighboring sources and zeroth order images. The GRISMR orientation has some off-center emission lines originating the neighboring source. This emphasizes the importance of having two orientations, as the GRISMC spectra confirms this emission belongs to a different source. 
    }
    \label{fig:linefinder}
\end{figure*}

The line-finding algorithm used in \citet{Battisti:2024} and \citet{revalski2024} has been updated to enable reliable analysis of the PASSAGE data:
the algorithm has been updated to handle up to three grism filters in up to two orientations, crucial for analyzing NIRISS grism data. 
Additionally, the algorithm now automatically generates region files that can be applied to both direct and dispersed images, to allow for an inspector to visually confirm the legitimacy of all identified emission lines and sources. The expected locations of bright zero-orders are also marked in the dispersed images. This feature is critical as zero-orders can be misidentified as emission features if they overlap with a trace. We note however that on one edge of the field (about $10\%$ of the field) we lack direct imaging to be able to mark the zero-orders of some off-field objects. % are not marked. 
As described in \cite{Battisti:2024}, the original software allows emission lines to be fit with Gaussian functions and the continuum is measured by interpolating between user-specifiable nodes using a cubic spline. Both of these functions are available in our updated version of the line fitting software, but we additionally allow emission lines to be fit with a double Gaussian containing both narrow and broad components. Moreover, the continuum can now be fit with three different functional families: splines, polynomials, and a linear fit. 
This added capability is necessary for the higher signal-to-noise JWST data (compared to HST data, on which the original software was based on) and allows for more reliable line flux and continuum measurements.

We require emission line candidates to consist of at least 3 contiguous pixels and a signal-to-noise for the emission line of $\rm S/N_{line}\geq 5$ \citep[see also][]{Colbert2013ApJ}. This corresponds to
$\rm S/N_{pixel}\geq 2.88$. 
The number of emission lines identified by the automatic line finding algorithm using a $5\sigma$ cutoff ranges from $\sim$250--1200 per field, depending on the exposure time and number of grism filters covered. 
This corresponds to $\sim$200--950 candidate emission line sources per field (for a total of 5176 candidates), as some objects contain multiple emission lines. Given that only a minority of \sep-detected sources are expected to show emission lines in the spectral range covered, the line-finding algorithm typically excludes $75\%$–$95\%$ of sources per field from further consideration, efficiently narrowing the candidate list for visual inspection.
In general, more emission lines are identified in NIRISS filters at longer wavelengths. The increase in detected lines from F115W to F150W to F200W arises for three main reasons. First, the wavelength coverage of the grism increases from 0.27,$\mu$m in F115W to 0.34,$\mu$m in F150W and 0.47,$\mu$m in F200W. Second, the two typically strongest rest-frame optical emission lines, \OIII\ and \Ha, are redshifted toward longer observed wavelengths, such that by $z\sim1$ \Ha\ falls outside the F115W bandpass, and by $z\sim1.6$ \OIII\ does as well.
Third, the number of artifacts that can masquerade as emission lines (such as zero order images) increases from the F115W to F200W.
Unsurprisingly, fields with three grism filter coverage (Par028 and Par052) produce the most candidate emission lines.

Following the automated line detection procedure, all emission lines are visually inspected to ensure their validity, and to confirm their redshifts.

\subsection{Visual Line Confirmation}\label{sec:VisualLineDetection}
The line-finding algorithm allows for visual inspection of every emission line object identified by the CWT algorithm. 
Figure~\ref{fig:linefinder} shows the visual-inspection interface for an example $z=1.87$ source in Par028. The top panel shows the 1D spectrum in black with the contamination model shown in pink. 
The grayed-out regions indicate the masked spectral ranges (between the F115W and F150W filters on the left and F150W and F200W on the right). 
The locations of prominent emission lines are marked with  vertical blue lines, and the corresponding emission lines are modeled with Gaussian profiles in red. The continuum is traced by the dashed blue lines, and the black points indicate the nodes used to constrain the continuum as a spline. 
The signal-to-noise ratio (SNR) at every wavelength is shown in the plot just below the 1D spectrum.

The direct images in the three filters are shown on the right, with the green circle indicating the target object. Grayscale images of the 2D spectra are seen in the bottom three rows, with each row corresponding to each NIRISS filter: F115W, F150W, and F200W. The two grism orientations are seen in the first and third columns, with the orientation-combined spectrum seen in the final column. Contamination modeling in the second and fourth columns indicate possible contamination for each orientation. 
For this object, the strong emission lines are seen in both orientations and cannot be attributed to contamination from neighboring sources, indicating these emission lines are real. 
The high SNR of \OII, \OIII, \Ha, and \Hb\ indicates this object has a very robust redshift. 
Emission lines that are marked on the visual line-finder interface to assist with line and redshift identification can be found in Table~\ref{tab:lines} in appendix~\ref{sec:emissionlinelist}.

\begin{table*}[]
    \centering
    \begin{tabular}{c|c|c}
         Flag & Case & Number of Sources\\
         \hline
         1 & Multiple emission lines with SNR$\geq5$ & 858 \\
         2 & One line with SNR$\geq5$ and at least one other line with $2\leq$SNR$<5$ & 425\\
         3 & Single \OIII\ blended doublet line (has signature asymmetry) with SNR$\geq5$ & 110\\
         4 & Other single line with SNR$\geq2$ or multiple lines with $2\leq$SNR$<5$ & 379\\
         1.5 & Originally flag 4 single-line emitters that have been adjusted/confirmed photometrically & 411\\
         \hline
    \end{tabular}
    \caption{Flags on PASSAGE redshifts scenarios and hierarchy. A lower flag indicates greater confidence in the spectroscopic redshift. The flag of 1.5 is discussed in section~\ref{sec:redshiftcomparison}. The third column displays the number of sources corresponding with each redshift flag. \label{tab:flag}}
\end{table*}

For every object in the field identified by the CWT, a (human) visual inspector must examine the direct image, the dispersed images, and 2D and 1D spectra. 
Only sources for which the CWT automated line-detection algorithm identifies at least one emission-line candidate are advanced to visual inspection. Each of the 15 fields is independently reviewed by two inspectors. For each source, a spectroscopic redshift is either assigned or the source is excluded from the final catalog if the object or its emission-line features are deemed spurious.

For a source to be assigned a spectroscopic redshift:
\begin{itemize}
\item Reviewers must verify the reality of each source using the imaging data. Sources are removed from the catalog at this stage if they are determined to be noise fluctuations, detector artifacts, or saturated stars.
\item Reviewers must also confirm that each emission line is genuine. Emission-line candidates are rejected at this stage if they do not appear in both grism orientations when both GR150R and GR150C data are available, if the feature is instead a poorly fit continuum segment (often due to overlap with a neighboring spectral trace), if the signal arises from a zeroth-order image overlapping the spectrum (which is marked in the line-finder), or if it is caused by a diffraction spike from a nearby star crossing the spectral trace, which can mimic an emission line in the one-dimensional spectrum.
\item Reviewers also verify that each emission line is correctly associated with the source and not attributable to a different object, as mis-association can lead to double counting or incorrect redshift assignments. In particular, the spatial morphology of the emission line should be consistent with that of the source, and the line feature should be centered on the spectral trace. An example of an off-trace contaminant is shown in the F115W GRISMR two-dimensional spectrum in Figure~\ref{fig:linefinder}.
\end{itemize}
If an object fails to pass any of these criteria, the object is removed from the catalog. 
Objects that pass the visual inspection criteria are assigned a spectroscopic redshift. Spectroscopic redshifts are primarily constrained using the following emission lines: \Ha, \OIII, \OII, \Hb, \SIII, \Pab, and \Paa. While other emission-lines are denoted in the visual interface, the aforementioned emission lines are typically the strongest and most reliable for redshift constraints.

For sources exhibiting multiple emission lines, the relative observed wavelengths must be consistent with the expected rest-frame line separations at a single spectroscopic redshift. The line-finding software allows users to assign a redshift, identify emission features as \Ha, \OIII, \OII, \Hb, \SIII, \Pab, or \Paa, or specify observed wavelengths at which individual lines are fit. A non-linear least-squares optimizer based on the Levenberg–Marquardt algorithm is then used to fit the full spectrum, including the continuum. In practice, contamination from neighboring sources—particularly in fields observed at only a single grism orientation—and low signal-to-noise emission lines can still lead to disagreements between reviewers, even for sources with multiple detected features.
An additional complication arises for sources with spatially offset emission, which may result from clumpy star formation, mergers, accretion, and/or outflows. Such off-center emission can produce multiple peaks within a single emission line and is discussed further in Appendix~\ref{sec:offcenter}.
For spectra containing a single emission line, the line is by default assumed to be \Ha\ and the corresponding redshift is assigned. However, if the line profile exhibits sufficient asymmetry to indicate the \OIII\ doublet, the feature is instead identified as \OIII\ and the redshift is adjusted accordingly.

\subsubsection{Reconciliation}\label{sec:recon}
After two reviewers have independently completed the visual inspection for each object in a field, there is a reconciliation procedure, in which both reviewers meet to discuss all objects in the field for which either there is a disagreement in the assigned redshift of the source,% (if $\Delta z>$0.2)
or if there is a disagreement in the validity of the source (i.e., one person removed the source from their catalog, and the other assigned it a redshift). 
The reviewers must come to an agreement on the redshift and legitimacy of the source to keep it in the final PASSAGE catalog.
In the rare case where no agreement is reached, the emission line is assigned the lowest data quality flag (see Section~\ref{sec:flag}), and the line is assigned as \Ha\ (or if there are multiple emission lines, the strongest is assigned as \Ha). The default assumption of \Ha\ is based on results from \citet{baronchelli2020}, which found that $>50\%$ of single line-emitters were correctly identified as \Ha\ in the absence of ancillary photometry.

\begin{table*}[]
    \centering
    \begin{tabular}{lll}
         Label & Description & Example Value\\
         \hline
         \textsc{id} & Source identifier & 1428\\
         \textsc{ra} & Right ascension [$^\circ$] & 150.085539\\
         \textsc{dec} & Declination [$^\circ$] & 2.43329\\
         \textsc{field} & PASSAGE field the object lies in & Par028\\
         \textsc{z\textunderscore best} & Best PASSAGE spectroscopic redshift & 1.8735\\   
         \textsc{emline\textunderscore flag} & Flag on the PASSAGE spectroscopic redshift & 1\\
         \textsc{z\textunderscore best\textunderscore err} & Best redshift error & 0.0000\\   
         \textsc{z\textunderscore passage\textunderscore linefinder} & Spectroscopic redshift identified via PASSAGE line-finding & 1.8735\\
         \textsc{z\textunderscore passage\textunderscore err} & Error on the PASSAGE spectroscopic redshift & 0.0000\\
         \textsc{cosmosweb\textunderscore id} & Source identifier in COSMOS2025/COSMOS-Web & 423415\\
         \textsc{cosmos2020\textunderscore id} & Source identifier in CLASSIC COSMOS2020 catalog & 1070736\\
         \textsc{confusion\textunderscore flag} & Confusion flag if multiple COSMOS objects could be matched &0\\
         & with a single PASSAGE object & \\
         \textsc{agn\textunderscore flag} & Flag if COSMOS-Web indicates the source is likely an AGN &0\\
         \textsc{warn\textunderscore flag} & COSMOS-Web warning flag &0\\
         \textsc{mass\textunderscore 50} & Stellar mass of the source (50th percentile), log$_{10}$[$\rm M_*/M_{\odot}$] & 8.601055\\ 
         \textsc{mass\textunderscore 84} & 84th percentile on stellar mass, log$_{10}$[$\rm M_*/M_{\odot}$] & 8.664838\\ 
         \textsc{mass\textunderscore 16} & 16th percentile on stellar mass, log$_{10}$[$\rm M_*/M_{\odot}$] & 8.532257\\ 
         \textsc{f115w\textunderscore niriss\textunderscore flux} & PASSAGE F115W auto flux [$\rm \mu$Jy] & 0.394123\\
         \textsc{f115w\textunderscore niriss\textunderscore err} & PASSAGE F115W auto flux error [$\rm \mu$Jy] & 0.005081\\
         \textsc{f150w\textunderscore niriss\textunderscore flux} & PASSAGE F150W auto flux [$\rm \mu$Jy] & 0.589730\\
         \textsc{f150w\textunderscore niriss\textunderscore err} & PASSAGE F150W auto flux error [$\rm \mu$Jy] & 0.005962\\
         \textsc{f200w\textunderscore niriss\textunderscore flux} & PASSAGE F200W auto flux [$\rm \mu$Jy] & 0.485601\\
         \textsc{f200w\textunderscore niriss\textunderscore err} & PASSAGE F200W auto flux error [$\rm \mu$Jy] & 0.005316\\
         \hline
    \end{tabular}
    \caption{Format of the redshift catalog. PASSAGE sources with no corresponding photometric match are given a -99 for any column requiring COSMOS data. The right ascension and declination are given in epoch J2000.\label{tab:catalog}}
\end{table*}

\subsubsection{Redshift Quality Flags}\label{sec:flag}

The line-finding and reconciliation process includes the assignment of data-quality flags to each source, reflecting the varying confidence in the derived spectroscopic redshifts. For example, redshifts based on multiple emission lines are generally more robust than those derived from a single line. The criteria for each redshift scenario and the corresponding quality flags are summarized in Table~\ref{tab:flag}.
Lower flag values indicate higher-confidence spectroscopic redshifts; the most secure classification (flag = 1) corresponds to sources with multiple emission lines detected at SNR$\geq5$. Only a subset of emission lines—\Ha, \OIII, \OII, \Hb, \SIII, \Pab, and \Paa—are used in defining this and the lower-confidence flags (2–4), as these lines provide the constraints on the redshift. 

A flag of 2 corresponds to sources with a single emission line detected at SNR$\geq5$ and at least one additional emission line with $2 \leq \mathrm{SNR} < 5$. Flags 1 and 2 therefore both denote multi-line emitters. We note that although the CWT automated line-detection stage requires emission-line candidates to have $\mathrm{S/N}_{\rm line} \geq 5$, lower-significance lines may still appear in the final catalog for two reasons: (1) multi-line emitters often exhibit additional weaker lines at the expected wavelengths and relative strengths, which nonetheless provide robust redshift confirmation; and (2) during visual inspection, reviewers may adjust the continuum fit to improve the overall spectral modeling, which can in some cases reduce the measured line SNR.

A flag of 3 is assigned when reviewers identify a single \OIII\ emission line with SNR$\geq5$. The characteristic asymmetry of the \OIII\ doublet, with an intrinsic flux ratio of $\sim$1:3 \citep{storey2000}, distinguishes it from other single-line features, resulting in a more reliable redshift than for other single-line emitters.

A flag of 4 is assigned to all remaining single–emission-line sources with SNR$\geq2$, as well as to sources with multiple emission lines for which all features have $2 \leq \mathrm{SNR} < 5$. For single-line flag~4 sources, the spectroscopic data alone provide limited redshift discrimination without photometric information; accordingly, approximately $80\%$ of these sources are provisionally identified as \Ha\ emitters during the line-finding process. %{\bf Section~\ref{sec:redshiftcomparison} describes photometric refinement of these flag~4 redshifts}. 
A small number of weak multi-line sources also fall into this category, for which the redshift assignments are likewise uncertain.

In Section~\ref{sec:redshiftcomparison}, we incorporate COSMOS photometric constraints for sources lying in the overlap between PASSAGE and COSMOS, which can in some cases refine or confirm the redshifts of flag~4 sources in hindsight. However, the line-finding procedure is intentionally performed without prior redshift information in order to assess the intrinsic reliability of the PASSAGE spectroscopic redshift assignments and to prepare for analyses of PASSAGE fields lacking ancillary photometric data.

We note that additional emission lines (e.g., \NeIII, \Pag, \SII, \OI, \HeI, \MgII, and higher-order Balmer lines such as \Hg\ and \Hd) may be present in some spectra but are typically too weak to be used reliably in the redshift flagging scheme. In cases where such lines are detected at high significance, the redshift is already robustly constrained by one or more of \Ha, \OIII, \OII, \Hb, \SIII, and/or \Pab.

A notable exception to the standard flag hierarchy is a highly secure $z=4.64$ quasar in Par028, which exhibits strong \OII, \NeIII, \MgII, and \CIII\ emission. Although this source would nominally be assigned flag~4 under the standard criteria, it is upgraded to flag~1 in the final catalog due to the unambiguous nature of its spectral features.

\section{Results}\label{sec:finalresults}
\subsection{Statistics of Rejected Sources}
The CWT automated line-detection algorithm initially identified 5,175 emission-line candidates within the COSMOS-Web/COSMOS2020 footprint. Following visual inspection (Section~\ref{sec:VisualLineDetection}), 2,183 sources were retained with assigned spectroscopic redshifts in the final PASSAGE catalog, while 2,992 candidates were rejected. Emission-line candidates were removed either because the source itself was determined to be spurious or because the identified emission line(s) were not deemed to be genuine.

On average, $53\%$ of emission-line candidates identified by the CWT automated detection were rejected during visual inspection, with a majority of candidates rejected in 7 of the 15 fields. This underscores the critical role of visual inspection: while the CWT algorithm substantially reduces the number of sources requiring review, it necessarily identifies a significant number of false-positive emission-line candidates. Common contaminants, such as stellar diffraction spikes, zeroth-order images, and overlapping spectral traces, can be readily recognized by human reviewers but remain challenging to identify reliably with automated methods alone.

Although a rejection rate of $\sim53\%$ may appear high, it is significantly lower than in comparable studies that rely on direct visual inspection of all \grizli-extracted sources. For example, \citet{noirot2023} identified 156 line emitters from 1394 sources, and \citet{watson2025} identified 488 line emitters from 3694 sources, corresponding to rejection rates exceeding $85\%$ in both cases. The use of the CWT algorithm therefore acts as an effective pre-filter, increasing the success rate of visual inspection to $\sim50\%$, depending on the field.

Raising the CWT detection threshold from $5\sigma$ to $10\sigma$ would further reduce the number of candidates requiring visual inspection, eliminating roughly $40\%$ of the sources identified at the $5\sigma$ level. However, this would also result in the loss of approximately $30\%$ of the sources ultimately accepted by the reviewers, substantially decreasing the size of the final emission-line sample.
For reference, reviewers reject approximately $65\%$ of candidates with at least one emission line detected at $\mathrm{SNR}>5$, compared to $\sim55\%$ of candidates with at least one line at $\mathrm{SNR}>10$. These trends further demonstrate that visual inspection remains essential for mitigating contamination and constructing a reliable spectroscopic sample.

\subsection{Spectroscopic Redshift Catalog}
\label{sec:catalog}

We present the PASSAGE spectroscopic redshift catalog in the COSMOS field, which can be found here: \url{https://github.com/huber458/Passage-in-Cosmos-Linelist-Catalog}. In total, the line-finding effort identifies 2,183 emission-line sources, corresponding to 3,208 emission lines detected at $\mathrm{SNR}>5$, as many sources exhibit multiple lines.\footnote{Due to the slight overlap between Par026 and Par028, one emission-line source appears in both fields. The redshift measured for this object is consistent across both fields; however, it is assigned a flag of 1 in Par028, where both \Ha\ and \SIII\ are detected, and a flag of 4 in Par026, where only \Ha\ is detected because of more limited wavelength coverage. To avoid double counting, we retain the Par028 measurements for this source.}

Each PASSAGE emission-line source is cross-matched with the COSMOS-Web and COSMOS2020 (CLASSIC) catalogs. A match is assigned to the nearest COSMOS source within a 0\farcs3 radius, chosen to correspond to the NIRCam F444W PSF FWHM, which is relevant for stellar mass measurements (Section~\ref{sec:stellarmasses}). If multiple COSMOS sources fall within this radius, the object is flagged as potentially confused (\textsc{confusion\textunderscore flag}) in the catalog. For cross-matched sources, the PASSAGE spectroscopic measurements are supplemented with multi-band COSMOS photometric data.

Of the 2,183 emission-line sources, 1,332 are matched to COSMOS-Web objects, 564 are matched to COSMOS2020 sources outside the COSMOS-Web footprint, and 287 have no counterpart in either COSMOS catalog. Based on the spectroscopic quality flags defined in Section~\ref{sec:flag}, 858 sources are assigned flag~1, 425 flag~2, 110 flag~3, and 790 flag~4; thus, the largest fraction of the catalog consists of the highest-confidence (flag~1) redshifts.

The catalog format is summarized in Table~\ref{tab:catalog}. It includes spectroscopic redshifts, quality flags, photometric fluxes, and stellar masses (Section~\ref{sec:stellarmasses}) for all PASSAGE emission-line sources detected in the 15 fields listed in Table~\ref{tab:obs_table}. The measured emission line fluxes and uncertainties will be presented in a forthcoming paper (Nedkova et al., \textit{in prep.}).
The preferred redshift and its associated uncertainty are provided in \textsc{z\textunderscore best} and \textsc{z\textunderscore best\textunderscore err}. COSMOS-Web and COSMOS2020 source identifiers are included for all cross-matched objects, while sources without COSMOS counterparts are retained in the catalog and assigned placeholder values of $-99$ for parameters requiring COSMOS photometric data.

Several quality and classification flags are included. The \textsc{emline\textunderscore flag} encodes the PASSAGE spectroscopic redshift confidence described in Section~\ref{sec:flag}. The \textsc{warn\textunderscore flag} mirrors the COSMOS-Web quality flag (indicating that a COSMOS-Web source may be a hot pixel, or that ground- and space-based photometry are inconsistent with one-another). The \textsc{agn\textunderscore flag} identifies sources classified as likely AGN in COSMOS-Web, as their photometry is better fit with an AGN SED template than a galaxy template. This flag indicates that the derived stellar masses in Section~\ref{sec:stellarmasses} may be unreliable.
%In COSMOS-Web , sources classified as potential AGN are done so if the SED photometric fit 

The column \textsc{z\textunderscore passage\textunderscore linefinder} records the redshift assigned during the line-finding procedure (with uncertainty \textsc{z\textunderscore passage\textunderscore err}) and matches \textsc{z\textunderscore best} for the vast majority of sources. For a subset of single-line emitters, the redshift is refined in Section~\ref{sec:redshiftcomparison} using COSMOS photometric information.

\begin{figure*}\label{fig:redshiftcomparison}
\centering
\includegraphics[width=1.4\columnwidth]{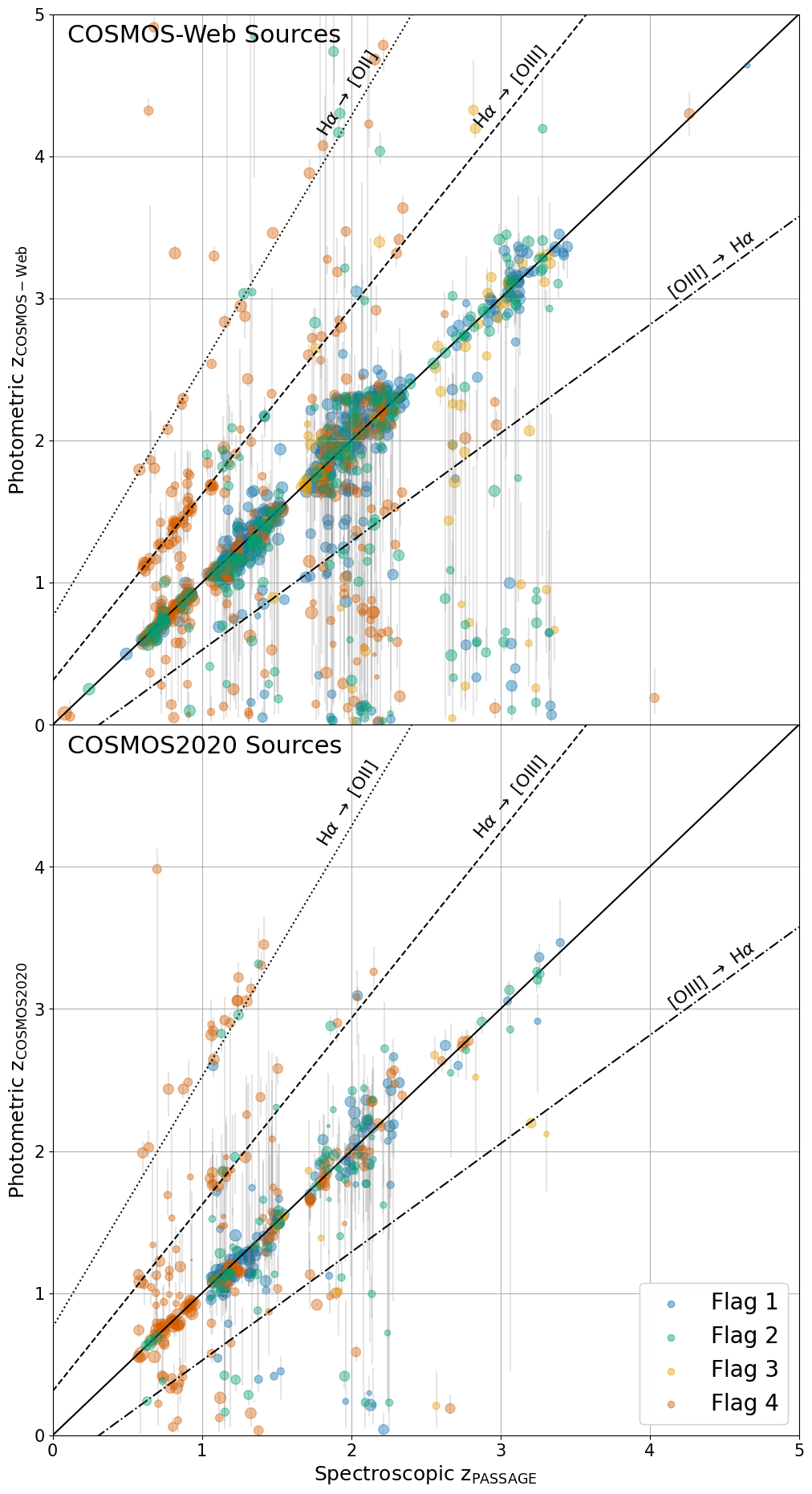}
\caption{Comparison of the photometric redshifts from COSMOS-Web (top panel) and COSMOS2020 (bottom panel) with the spectroscopic redshift from the PASSAGE line-finding efforts. The color indicates the quality flag on the redshift. The ACS/F814W magnitude of the sources is indicated by the size of the marker: smaller markers indicate a fainter magnitude. The solid line represents an exact match in redshift. The dashed (dotted) line traces the redshifts where the PASSAGE team would have identified a line as \Ha, but the corresponding redshift from COSMOS-Web would have identified it as \OIII (\OII). The dash-dotted line traces the redshifts where the PASSAGE team would have identified a line as \OIII, but the corresponding redshift from COSMOS-Web would have identified it as \Ha.}
\end{figure*}

\section{Redshift Comparison}\label{sec:discussion}
The overlap between PASSAGE and COSMOS provides both spectroscopic and photometric redshift measurements for a large sample of sources. We begin by comparing the redshifts derived from these two catalogs.

\subsection{Photometric Redshift Comparison}\label{sec:redshiftcomparison}
The availability of both PASSAGE and COSMOS data enables consistency checks between independent redshift measurements.
COSMOS-Web photometric redshifts have been shown to be fairly accurate: simulations conducted in \citet{casey2023} find that the normalized median absolute deviation, $\sigma_{NMAD}$ (a measure of variation that suppresses outliers), defined as 
\begin{equation}\label{eq:nmad}
\sigma_{NMAD}=1.48\times \rm median \bigg( \bigg| \frac{\Delta z}{1+z} - median \bigg(\frac{\Delta z}{1+z} \bigg) \bigg| \bigg)
\end{equation}
following \citet{hoaglin1983,brammer2008,casey2023,wang2024,ratajczak2025}, where 
\begin{equation}
\frac{\Delta z}{1+z} = \frac{z_{phot}-z_{spec}}{1+z_{spec}}.
\end{equation}
between photometric redshifts fit to mock galaxies using the full suite of COSMOS-Web filters and the true spectroscopic redshifts to be $\sigma_{NMAD}=0.033$. This indicates a high degree of redshift accuracy from the COSMOS-Web photometry. 
However, without spectroscopic information, these photometric redshifts are often accompanied by large error bars and contain sources with catastrophic redshift errors \citep[e.g.,][]{casey2023, ratajczak2025}. 
The PASSAGE spectroscopic redshifts have very low uncertainties (typically on the order of 0.0005), assuming that all emission lines in PASSAGE spectra have been identified correctly. Realistically, not all emission lines in PASSAGE will have been identified correctly \citep[see][]{baronchelli2020,baronchelli2021,boyett2024}, and thus this assumption is only true for a subset of galaxies. The question is: for which subset of sources is this true?
For PASSAGE flag 1 multi-line emitters, the identity of emission lines is extremely reliable, as it is difficult to find distributions of strong emission lines that line up perfectly with observed PASSAGE lines other than their true identity, and especially not with reasonable flux ratios.
On the other hand, the PASSAGE redshifts for flag 4 single-line emitters can still precisely constrain the peak of the single emission line, but the lack of other lines in the wavelength range means that the identity of the single-emission line is often ambiguous. 
In these instances, the COSMOS photometry can often provide some insight into the true identity of the single-line emitter.

In the top panel of Figure~\ref{fig:redshiftcomparison}, we compare COSMOS-Web cross-matched PASSAGE spectroscopic redshifts with the COSMOS-Web photometric redshifts. In the bottom panel, we do the same, but for COSMOS2020 cross-matched sources that lie outside of COSMOS-Web.
The color of the points indicates the flag on the redshift. 
The ACS/F814W magnitude of the sources is indicated by the size of the marker: smaller markers indicate a fainter magnitude.
The solid black line represents the 1:1 line. 
For the combined COSMOS-Web/COSMOS2020 sample, $65\%$ of redshifts agree within $|\Delta z|=|z_{phot}-z_{spec}|\leq 0.2$, and $78\%$ of redshifts agree within a $|\Delta z|\leq 0.5$. This indicates that in the vast majority of cases, $z_{PASSAGE}$ and $z_{COSMOS}$ are in relatively good agreement. This agreement does depend on the redshift flag: we find that for the subset of sources with a quality flag of 1 that $\sigma_{NMAD}=0.036$. 
This indicates a very strong agreement between the photometric and spectroscopic redshifts, and essentially mirrors the level of agreement in the aforementioned \citet{casey2023} simulations between photometric and spectroscopic redshifts in COSMOS-Web. 
That said, for the flags 2 and 3 subsets, $\sigma_{NMAD}$ rises to $\sigma_{NMAD}=0.071$ %indicating a weaker redshift agreement. 
and $\sigma_{NMAD}=0.109$ respectively, indicating an increasingly weaker redshift agreement.
For the flag 4 subset, the $\sigma_{NMAD}=0.094$, indicating a relatively weak redshift agreement, but there is an overabundance of flag 4 sources located near the dashed and dotted black lines in Figure~\ref{fig:redshiftcomparison}: the dashed line represents the locations where the PASSAGE reviewers would classify a single-emission line as \Ha (our conservative default choice), but the COSMOS photometry suggests that the emission line is actually \OIII. Likewise, the dotted line represents the galaxies where the PASSAGE reviewers would classify a single-emission line as \Ha, but the COSMOS photometry suggests that the emission line is actually \OII. Therefore, with the benefit of hindsight, it is likely that many of the sources that lie close to these lines are mis-identified ambiguous single-line emitters. 

For a subset of flag~4 single-line sources initially identified as \Ha\ emitters, we refine the PASSAGE spectroscopic redshifts using COSMOS photometric redshift information. This refinement assumes that, in some cases, the single detected emission line was misidentified as \Ha\ and instead corresponds to \OIII\ (dashed-line scenario) or \OII\ (dotted-line scenario). Because the observed wavelength of the emission feature is well constrained by PASSAGE, directly adopting the COSMOS photometric redshift would be inappropriate. Instead, we compute an adjusted redshift using
\begin{equation}
z_{\rm best} = \frac{\lambda_{\rm H\alpha,0}}{\lambda_{\rm OIII,0}} (1+z_{\rm spec}) - 1
\end{equation}
for the \OIII\ case, and
\begin{equation}
z_{\rm best} = \frac{\lambda_{\rm H\alpha,0}}{\lambda_{\rm OII,0}} (1+z_{\rm spec}) - 1
\end{equation}
for the \OII\ case.

This refinement is applied to all single-line \Ha-designated sources that lie within $5\sigma_{\rm NMAD}$ of the dashed or dotted loci in Figure~\ref{fig:redshiftcomparison}, where $\sigma_{\rm NMAD}=0.037$ is derived from the robust flag~1 sample. This threshold corresponds approximately to sources within $\Delta z \simeq 0.18$ of the expected loci.

Sources whose redshifts are refined in this manner will show the refined redshift in the \textsc{zbest} column of the catalog (but the original redshift from the lin-finding effort can still be found in the \textsc{z\textunderscore passage\textunderscore linefinder} column). These sources are also reclassified as flag~1.5, together with flag~4 sources that already show strong agreement between $z_{\rm PASSAGE}$ and $z_{\rm COSMOS}$ within the same $5\sigma_{\rm NMAD}$ threshold. This designation reflects that these redshifts are substantially more reliable than those of the remaining flag~4 sources.
After this procedure, 183 flag~4 sources with COSMOS counterparts remain discrepant, with $\sigma_{\rm NMAD}=0.400$, and an additional 224 flag~4 sources lack COSMOS counterparts entirely. The PASSAGE spectroscopic redshifts for these 407 sources should therefore be treated with caution. In contrast, 383 sources are classified as flag~1.5, exhibiting $\sigma_{\rm NMAD}=0.022$, and their redshifts can be considered robust.
 
Although photometric redshifts are themselves subject to uncertainties, the accuracy demonstrated in simulations by \citet{casey2023}, combined with the fact that nearly all sources with $z_{\rm COSMOS} > z_{\rm PASSAGE}$ align closely with the expected \OIII\ and \OII\ loci, suggests that this adjustment provides the most plausible redshift estimates for the majority of flag~1.5 sources. In total, 81 \OIII\ lines and 27 \OII\ lines are re-identified through this process, corresponding to approximately $14\%$ and $5\%$, respectively, of all cross-matched flag~4 sources. Thus, with the benefit of photometric hindsight, we estimate that $\sim19\%$ of flag~4 single-line sources were initially misidentified.

The number of photometrically adjusted redshifts varies by field configuration. The five single-filter fields account for 33 adjusted sources, the eight two-filter fields for 65, and the two three-filter fields for 10. This distribution is expected, as fields with more limited wavelength coverage naturally produce a larger fraction of flag~4 sources. Notably, Par051 alone contributes 20 adjusted redshifts, nearly one-fifth of the total. Unlike the other single-filter fields, which were observed in F200W, Par051 was observed only in F115W, where the lower spectral resolution makes the characteristic asymmetry of the \OIII\ doublet more difficult to identify during visual inspection. As a result, sources that would otherwise be classified as flag~3 are more likely to be classified as flag~4 in this field, consistent with Par051 also having the fewest flag~3 sources of any field.

One consequence of this photometric refinement is that many of the highest-redshift sources in the catalog ($3<z<5$) are identified through this adjustment, primarily as single \OII\ emitters detected in F200W, which are otherwise difficult to distinguish from \Ha\ without photometric context. This has important implications for the remaining PASSAGE fields outside COSMOS, as well as for PASSAGE sources without COSMOS counterparts: in the absence of ancillary photometric data, flag~4 single-line sources are likely to be misidentified at the $\sim19\%$ level.

An additional overdensity is visible in Figure~\ref{fig:redshiftcomparison} along the dash–dotted line. This locus corresponds to cases where COSMOS photometric redshifts would favor an interpretation of an emission feature identified by PASSAGE as \OIII\ instead of being \Ha\ by the reviewers. However, the majority of sources near this locus are multi-line emitters rather than flag~3 single-line sources, and renewed visual inspection confirms the robustness of the PASSAGE redshift assignments.

A likely contributor to this overdensity is the absence of F200W imaging in COSMOS-Web. While JWST NIRCam provides imaging in F115W, F150W, F277W, and F444W, it does not include F200W coverage. A common configuration for sources in this region of parameter space is the presence of \OIII\ emission in F150W and \Ha\ emission in F200W. Although the PASSAGE spectroscopic redshifts for these sources are well constrained, the lack of F200W imaging in COSMOS can lead SED fitting codes to misinterpret the strong F150W flux as originating from \Ha\ rather than \OIII.

There are a few reasons why sources with robust PASSAGE spectroscopic redshifts are significantly different from the COSMOS-Web redshifts. Many COSMOS-Web sources have large error bars, so that even if the best value is discrepant, the redshifts are still consistent within $1\sigma$. Some sources have a poor fit to the photometry, as indicated by a high $\chi^2$ value in the \citet{shuntov2025} catalog. Some sources have multiple peaks in the P(z) that are not well reflected by the error bars, indicating that several redshift ranges can fit the observed photometry. Other sources are among the faintest objects in the sample, for which redshift disagreements are not unexpected.

\begin{figure}%\label{fig:redshiftspeccomp}
\includegraphics[width=1.\columnwidth]{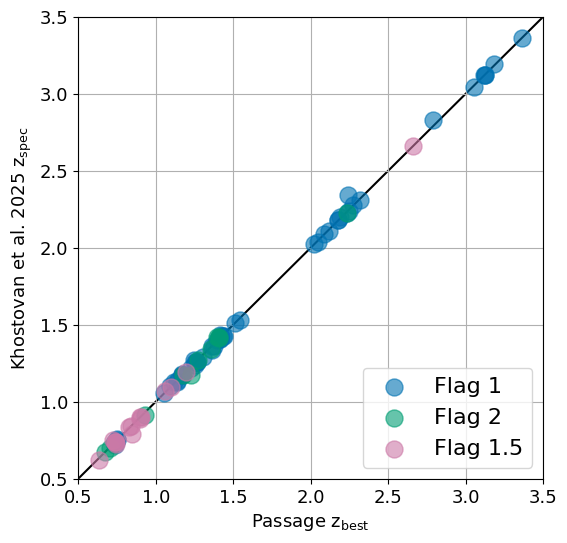}
\caption{Spectroscopic redshift comparison between the PASSAGE redshifts and the ancillary spectroscopic redshifts compiled in \citet{khostovan2025}. Only high quality spectroscopic redshifts from \citet{khostovan2025} ($\rm Q_f =3, 4$) and PASSAGE (flag 1, 2, and 1.5) are included in this comparison. \label{fig:speczcomp}}
\end{figure}

\subsection{Spectroscopic Redshift Comparison}\label{sec:specredshiftcomp}

We now compare the PASSAGE spectroscopic redshifts to previously obtained spectroscopic redshifts. 
\citet{khostovan2025} recently presented a compilation of the COSMOS spectroscopic redshifts encompassing $\sim20$ years of spectroscopic
redshifts within a $\sim10$ deg$^2$
area centered on the COSMOS legacy field. Using the same cross-matching parameters as in section~\ref{sec:catalog}, we cross-match the PASSAGE catalog with the \citet{khostovan2025} compilation. 
There are 228 cross-matched sources between catalogs. 
We restrict our comparison to the 81 out of 228 sources that have robust redshifts in both samples (the sources that have a quality flag of $\rm Q_{f}=3$ or $4$ in the \citet{khostovan2025} sample and have a flag of 1, 2, or 1.5 in the PASSAGE sample). In the \citet{khostovan2025} sample, a source with $\rm Q_{f}=3$ or $4$ has a $>95\%$ confidence in the redshift.
See appendix~\ref{sec:khostcomp} for a comparison with lower quality flags. In figure~\ref{fig:speczcomp}, we show the comparison between the \citet{khostovan2025} spectroscopic redshifts and the PASSAGE spectroscopic redshifts. All 81 cross-matched sources have a redshifts agreement between catalogs at the level of $|z_{Khostovan+}-z_{best}|<0.1$.

This comparison also emphasizes how powerful PASSAGE (and in general NIRISS) is in providing a legacy trove of spectroscopic redshifts: the 15 $\sim2.2'\times2.2'$ PASSAGE fields provide 2183 spectroscopic redshifts, of which only 228 sources had existing spectroscopic redshifts (of various redshift robustness). This means that PASSAGE has delivered 1955 spectroscopic redshifts in COSMOS for sources that previously lacked any such measurement, and demonstrates the strength of the pure-parallel approach with NIRISS.

\subsection{Redshift Comparison to \grizli}
\begin{figure}
\includegraphics[width=1.\columnwidth]{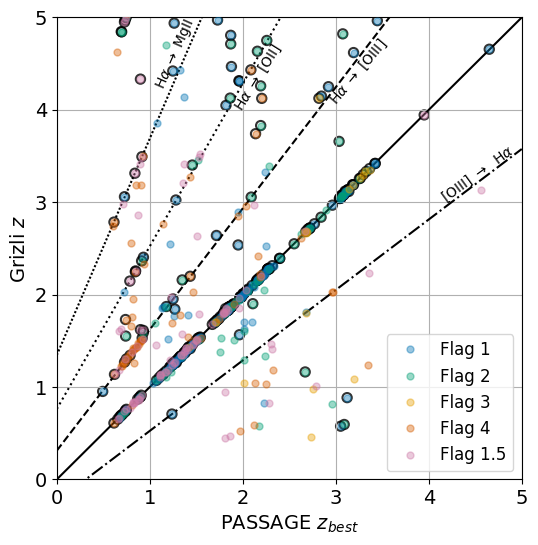}
\caption{Spectroscopic redshift comparison between the $z_{best}$ PASSAGE redshifts and the \grizli\ derived spectroscopic redshifts for four PASSAGE fields. Sources from Par028 are circled with a black outline. \label{fig:grizlicomp}}
\end{figure}

It is also worth exploring how the redshift obtained from the line-finding procedure compare with those from blindly running the \grizli\ algorithm to determine the redshifts. For Par003 (F200W only), Par025 (F115W and F150W), Par028 (F115W, F150W, and F200W), and Par051 (F115W only), which represent each combination of NIRISS filters in this study, we compare the \grizli\ predicted redshifts to $z_{best}$. This comparison is seen in figure~\ref{fig:grizlicomp}, and does not include sources that were rejected by the reviewers (of which \grizli\ would have blindly assigned a redshift). 

Just $65\%$ of the sources in this sample have a redshift agreement of $|z_{\grizli}-z_{best}|<0.1$. 
Flag 1 sources tend to agree in redshift ($85\%$ of flag 1 sources have $|z_{\grizli}-z_{best}|<0.1$), but this discrepancy drops with increasing flag (down to $52\%$ for flag 2 sources). 
However, even in the flag 1 subset, there are noticeable outliers. For example, there is a flag 1 object in Par028 with a $z_{best}=1.82$ and $z_{\grizli}=4.04$. The $z_{best}=1.82$ redshift assignment comes from \Ha, \OIII, and \Hb\ lines that are seen in both grism orientations. However, the $z_{\grizli}=4.04$ would assign \MgII\ to only the \OIII\ peak and does not account for the legitimate \Ha peak. There are a number of single-line emitters that are assigned \MgII\ by \grizli\ (see sources near the densely-dotted line in figure~\ref{fig:grizlicomp}) but the line-finding procedure (and in some cases the COSMOS photometry) would have assigned these lines to be \Ha. 

For the flag 1.5 sources where the line-finding $z_{spec}$ agrees with $z_{COSMOS}$ (ignoring for the moment subset of flag 1.5 sources with photometrically 'adjusted' redshifts), only $33\%$ of \grizli\ redshifts agree within $\Delta z\leq 0.1$ of the other two measurements of redshift (which are in agreement with each other). This means that for this selection of single-line emitters, \grizli\ is actually more likely to disagree than agree with $z_{best}$. 

If we cull the comparison to just Par028 (the field with 3 filters, indicated the points with black circles in Figure~\ref{fig:grizlicomp}), $75\%$ of redshifts agree between the \grizli\ $z$ and $z_{best}$. This agreement rises to $88\%$ for just the Par028 flag 1 sources. This indicates that the primary disagreement between the two measurements of redshift is from sources that have just one emission line in the wavelength coverage. A source of disagreement for flag 1 multi-line emitters is due to the similar spacing between different pairs of strong emission lines (e.g., the lines of \Ha\ and \Hb\ are spaced apart very similarly to the lines of \OIII\ and \OII). 

We compare these \grizli\ redshifts directly with the COSMOS-Web redshifts. The $\sigma_{NMAD}=0.082$ is larger than the corresponding comparison between the line-finding redshifts (prior to any photometric redshift adjustments) and the COSMOS-Web redshifts, which has a $\sigma_{NMAD}=0.062$. 
This comparison to \grizli\ does not include any of the sources that were rejected by the reviewers, meaning \grizli\ would also provide additional redshifts for spurious sources.
This does not mean that \grizli\ is for sure incorrect in any cases were it disagrees with both COSMOS-Web and the line-finding procedure, but it is a strong indicator that \grizli\ is typically less accurate the line-finding visual inspection procedure. Hence the line-finding effort is required to get the most accurate spectroscopic redshifts.

\section{Redshift Statistics}\label{sec:redstats}
\begin{figure}\label{fig:redshifthistogram}
\includegraphics[width=1.\columnwidth]{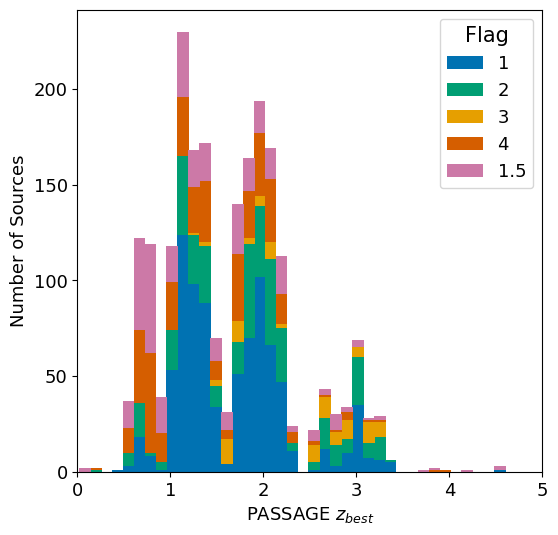}
\caption{Histogram of the finalized PASSAGE spectroscopic redshifts, $z_{best}$. The stacked histogram is broken up by flag. We identify sources covering a redshift range of $0.08\lesssim z_{spec} \lesssim 4.7$.}
\end{figure}

We now proceed with the full catalog, including the newly defined flag~1.5 sources, and examine its statistical properties, beginning with the redshift distribution. Figure~\ref{fig:redshifthistogram} shows a stacked histogram of the spectroscopic redshifts in the COSMOS PASSAGE catalog, with each bin subdivided by redshift quality flag.
The three prominent gaps in the redshift distribution at $z\sim1$, $z\sim1.6$, and $z\sim2.4$ correspond to redshift ranges where \Ha, \OIII, and \OII\ fall into the $\sim13{,}000$,\AA\ and $\sim17{,}000$,\AA\ gaps between the F115W/F150W and F150W/F200W filter bandpasses, respectively (e.g., see Figure~\ref{fig:linefinder}).

Galaxies with high-confidence flags (1 and 2) lie primarily in the $1.0\lesssim z_{spec} \lesssim 3.4$ range. This is because for multi-filter fields, \Ha\ and \OIII\ can appear in neighboring NIRISS filters at $1.0\lesssim z_{spec} \lesssim 2.3$, and \OIII\ and \OII\ can appear in neighboring NIRISS filters together at $1.7\lesssim z_{spec} \lesssim 3.4$. 
Flag~3 sources become more prevalent at higher redshifts. This flag is most commonly assigned in single-filter fields, where, as discussed in Section~\ref{sec:redshiftcomparison}, the characteristic asymmetry of the \OIII\ doublet is more readily identified in the F200W grism. \OIII\ can be detected in F200W over the redshift range $2.4 \lesssim z_{\rm spec} \lesssim 3.4$. At $z \gtrsim 3.5$, redshifts are most often assigned when \OII\ falls within the F200W bandpass, which in many cases is only achievable through the photometric redshift refinement unless additional features, such as \NeIII\ or higher-order Balmer lines, are also detected. A small number of sources at these redshifts exhibit \MgII\ emission, potentially indicating the presence of an AGN.

At the low-z end, line-identification is dominated by \SIII\ and \Pab\ lines. For a source to have $z\lesssim0.5$, \SIII\ must be identified in the F115W or F150W and/or \Pab\ must be identified in the F150W or F200W. 
Single-line \Ha\ emitters are often identified alone as a flag 4 or flag 1.5 source at $0.5\lesssim z\lesssim 1$ in the F115W, at $1\lesssim z\lesssim 1.7$ in the F150W  and at $1.7\lesssim z\lesssim 2.2$ in the F200W. 
The redshift range $1.7 \lesssim z \lesssim 2.2$ contains some of the most robust redshift measurements, as in three-filter fields \Ha, \OIII, \Hb, and \OII\ all fall within the NIRISS wavelength coverage. This makes this interval particularly well suited for metallicity studies and dust corrections using the Balmer decrement.

\section{Stellar Masses}\label{sec:stellarmasses}
The stellar masses of emission-line sources cross-matched between PASSAGE and COSMOS are derived using the COSMOS-Web photometry, with updated redshifts and additional PASSAGE photometry in the NIRISS F115W, F150W, and F200W filters where available. 
Owing to the differences in the data reduction and method adopted for the photometric extraction, we use here only photometry from the COSMOS-Web DR1 release \citep{shuntov2025} rather than combining with the COSMOS2020 data.
We use the \texttt{SourceXtractor++} \citep[hereinafter SE++]{bertin2020,kummel2022} fluxes in all available bands, combined with the PASSAGE NIRISS \texttt{auto} fluxes (see Table~\ref{tab:catalog}).
In the case where multiple sources in PASSAGE match to the same source in COSMOS-Web, due to the de-blending parameters chosen in Section~\ref{sec:passagedatareduction}, we rescale the COSMOS-Web fluxes for each source by the ratio of their PASSAGE \texttt{flux\_auto} measurements.

As the NIRISS filters originated as ``flight spares'' for NIRCam \citep{doyon2023}, they have very similar transmission curves.
This enables us to check for any systematic biases in our flux measurements (e.g. from differing background subtraction, photometric calibrations, or segmentation) using the PASSAGE NIRISS and COSMOS-Web NIRCam fluxes in the F115W and F150W filters.
We measure the fractional flux difference, $(f_{\rm{COSMOS-Web}}-f_{\rm{PASSAGE}})/f_{\rm{COSMOS-Web}}$, for cross-matched sources in F115W (F150W), finding a median offset of 1.5\% (-0.7\%) with scatter $\sigma=22\%$ (18\%).
The scatter is higher than that observed in previous studies \citep[e.g.][]{watson2025a}, consistent with the lower median S/N in our catalog, but we are satisfied that there are no systematic variations between the two sources of photometry.

Stellar masses were derived using a similar procedure to that described in \citet{Malkan.2025}, summarized here.
We model the stellar populations of each galaxy by fitting the
photometry with the Bayesian Analysis of Galaxies for Physical Inference and Parameter EStimation \citep[\textsc{Bagpipes}\footnote{\url{https://bagpipes.readthedocs.io}, version 1.3.2}; ][]{carnall2018} SED fitting code, using a 7-bin continuity SFH prior \citep{leja2019}. 
We fix the youngest age bin to 30\,Myr, and the oldest to 500\,Myr below the age of the universe at a given redshift, permitting both a maximally old and young population.
We allow the metallicity and ionization parameter to vary between (0.0,3.0)\,$Z_{\odot}$ and (-3.5,-1.0) respectively. Dust attenuation was parameterized following \citet{calzetti2000}, allowing an $A_V$ of up to three magnitudes. The masses for AGN will be highly unreliable, and we include an AGN flag in the catalog for suspect AGN (as identified in the COSMOS-Web DR1 release).

\begin{figure*}
    \centering
    \includegraphics[width=1\linewidth]{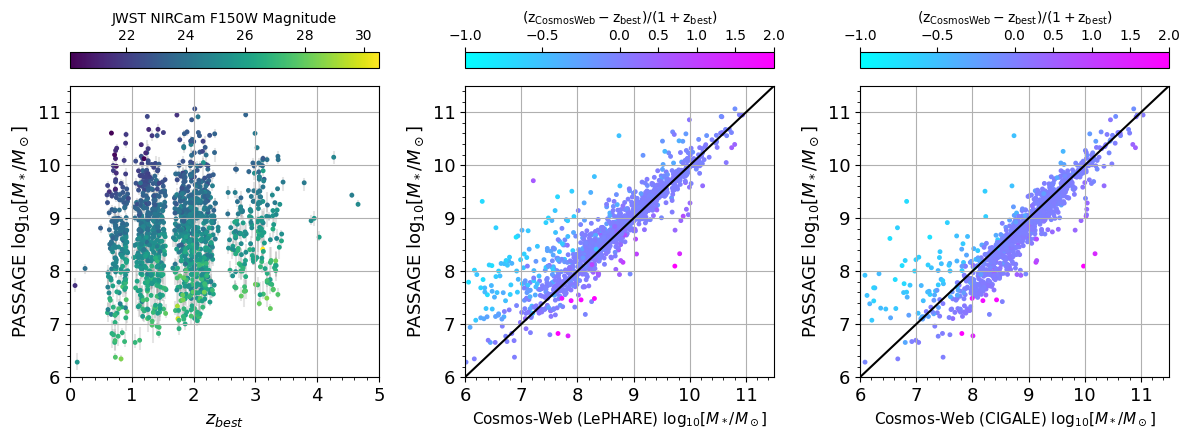}
    \caption{Left panel: stellar mass - redshift comparison for the PASSAGE galaxies. The lower mass limit on PASSAGE observations increases with redshift, as the least massive galaxies are harder to observe with increasing distance. Middle panel: stellar mass comparison between the COSMOS-Web LePHARE derived massed and the PASSAGE derived masses. The color-bar indicates the discrepancy between the COSMOS-Web and PASSAGE redshifts. PASSAGE tends to predicts a larger mass than COSMOS-Web LePHARE masses, which correlates with a larger measured PASSAGE redshift. Right panel: Same comparison as the middle panel, except with COSMOS-Web CIGALE derived masses instead of the LePHARE ones on the x-axis. PASSAGE tends to predicts a smaller mass than COSMOS-Web CIGALE masses.
    Error bars on the middle and right panels are omitted for legibility.}
    \label{fig:masses}
\end{figure*}

In the left panel of Figure~\ref{fig:masses}, we present the stellar masses at different redshifts.
%we compare the PASSAGE stellar masses to the corresponding PASSAGE redshifts. 
Stellar masses are observed to be between $10^6$ and $10^{11} \rm M_\odot$.
The color of each point corresponds to the JWST NIRCam F150W magnitude.
The low-mass limit increases with redshift due to the reduced apparent brightness of distant sources. We observe sources below $10^{7} \rm M_\odot$ galaxies as far as $z\sim1.5$ and below $10^{8} \rm M_\odot$ galaxies as far as $z\sim3.3$, substantially exploring the mid- and low-mass galaxy population near and past cosmic noon. 

In the middle panel of Figure~\ref{fig:masses}, we compare the PASSAGE derived stellar masses (which includes the COSMOS-Web photometry and additional NIRISS photometry and fixed to the PASSAGE $z_{best}$) to the LePHARE SED fitting code \citep{arnouts2002,ilbert2006} derived masses in the COSMOS-Web DR1 release.
The colorbar on this panel indicates the redshift discrepancy between PASSAGE and COSMOS-Web.
We find that the PASSAGE masses on average are measured to be larger than the LePHARE COSMOS-Web masses by $\sim$0.2 dex. In the right panel of Figure~\ref{fig:masses}, we compare the PASSAGE derived stellar masses to the CIGALE SED fitting code \citep{boquien2019} derived masses in the COSMOS-Web DR1 release. In contrast with the LePHARE derived masses, the PASSAGE masses are typically $\sim0.1$ dex lower than the COSMOS-Web measured masses using the CIGALE SED fitting code. In both the LePHARE and CIGALE comparisons, the typical mass discrepancy is smaller than the scatter (the root mean squared error between the COSMOS-Web and PASSAGE masses is $\sim0.3$ dex in the LePHARE case and $\sim0.5$ dex in the CIGALE case).
The biggest factor driving mass discrepancy is redshift discrepancy: when COSMOS-Web predicts a smaller redshift for a source than PASSAGE, it also predicts a smaller mass.
The scatter between PASSAGE and COSMOS-Web masses is typically more significant on the lower mass end: for sources with COSMOS-Web $\rm log_{10}[M_*/M_\odot] \lesssim7$ (for both the LePHARE and CIGALE cases), the corresponding PASSAGE stellar masses are typically larger by $\sim1.4$ dex. This low-mass discrepancy is primarily driven by the fact that these objects also typically have the most discrepant redshifts (with PASSAGE predicting a higher redshift than COSMOS-Web). 

To test the effect of both the addition of the NIRISS filters, we re-run SED fitting on Par028 without the NIRISS filters, but with the redshift still fixed to $z_{best}$. We find that the root mean squared error between the Par028 PASSAGE masses with and without the addition of the NIRISS filters to be 0.03 dex. For reference, we also re-run the SED fitting on Par028 with the NIRISS filters, but with $z_{COSMOSWeb}$ instead of $z_{best}$. This gives a root mean squared error of 0.38 dex, indicating redshift discrepancy more substantially influences the stellar masses than the additional photometry does.

\section{Conclusion}\label{sec:conclusion}
We present a spectroscopic redshift catalog of 2,183 galaxies spanning $0.08 < z < 4.7$ from 15 PASSAGE fields in the COSMOS region. We describe the emission-line identification pipeline and visual inspection procedure, which together mitigate contamination and assign reliability flags to each source, with lower flag values indicating higher-confidence redshifts.

Of the 2,183 emission-line sources, 1,955 previously lacked spectroscopic redshifts. Cross-matching with COSMOS photometric catalogs shows strong agreement between PASSAGE spectroscopic redshifts and photometric redshifts for the most robust (flag~1) sources, with increasing scatter for higher flag values. For the lowest-confidence redshifts (flag~4), we find that single-line \Ha\ emitters are misidentified at the $\sim19\%$ level. Using COSMOS photometric information, we confirm or refine the redshifts of a subset of these sources. We also find excellent agreement between PASSAGE redshifts and existing spectroscopic measurements from the \citet{khostovan2025} compilation. 

For sources within the COSMOS-Web footprint, we derive stellar masses and show that PASSAGE probes the low- and intermediate-mass galaxy population, reaching $\log_{10}(M_*/M_\odot) \lesssim 8$ out to $z \sim 3.3$. PASSAGE-derived stellar masses are systematically offset relative to COSMOS-Web estimates derived with \textsc{LePhare} and \textsc{CIGALE}, with differences primarily driven by redshift discrepancies. 

The methodology and results presented here have important implications for the broader PASSAGE survey. The remaining 48 PASSAGE fields lack uniform ancillary photometry, and the limitations identified for spectroscopic redshifts—particularly for single-line emitters—must be accounted for in future analyses. Ongoing follow-up efforts, including HST-SNAP, VLT/FORS2, and Keck observations, will provide critical photometric constraints to improve redshift assignments and enable stellar mass measurements in fields without existing ancillary data. More generally, this work highlights the value of incorporating photometric information when interpreting single-line slitless spectra, a lesson applicable to other pure-parallel surveys such as POPPIES \citep{poppies2024} and SAPPHIRES \citep{egami2024}.

Finally, this catalog provides an important reference dataset for upcoming Euclid and Roman spectroscopy, both of which will observe the COSMOS field. The PASSAGE redshift sample will enable cross-comparisons and consistency checks with Euclid and Roman measurements, while future spectroscopy from these missions will further test and refine PASSAGE redshifts, particularly for lower-confidence sources. Such cross-validation will be especially valuable given that approximately $44\%$ of the full PASSAGE survey consists of fields with single-filter spectroscopy.

\begin{acknowledgements}
This research was supported by the International Space Science Institute (ISSI) in Bern, through ISSI International Team project \#24-624. This work was supported by NASA through grant no. JWST-GO-1571. All the JWST data used in this paper can be found in MAST: \dataset[10.17909/6ca5-ba17]{http://dx.doi.org/10.17909/6ca5-ba17}.
AA, BV, and PW acknowledge support from the European Union – NextGenerationEU RFF M4C2 1.1 PRIN 2022 project 2022ZSL4BL INSIGHT. PW and BV acknowledge support from the INAF Mini Grant ``1.05.24.07.01 RSN1: Spatially-Resolved Near-IR Emission of Intermediate-Redshift Jellyfish Galaxies'' (PI Watson). AJB acknowledges funding from the “FirstGalaxies” Advanced Grant from the European Research Council (ERC) under the European Union’s Horizon 2020 research and innovation program (Grant agreement No. 789056).

\software{NumPy \citep{Numpy:2020}; SciPy \citep{Scipy:2020}; AstroPy \citep{Astropy:2013,Astropy:2018,Astropy:2022}; Matplotlib \citep{Matplotlib:2007}; \grizli\ \citep{Grizli:2021}; \sep\ \citep{barbary2016,bertin1996}.}

\end{acknowledgements}

%\facility{JWST}

\clearpage

\begin{appendix}
%\newpage

\section{Emission Lines in Line-Finder}\label{sec:emissionlinelist}
Table~\ref{tab:lines} lists all emission lines that are included in the line-finder to assist with visual identification.

\begin{table}[h]
\centering
\begin{tabular}{c|cccc}
\textbf{Emission Line}                         & \textbf{Vacuum Wavelength (\AA)} & \textbf{} & \textbf{} &  \\ \cline{1-2}
\Lya                   & 1215.67                                   & \textbf{} & \textbf{} &  \\
\Nv                             & 1238.82, 1242.80                                      &           &           &  \\
\CIV                            & 1548.20, 1550.78                                      &           &           &  \\
\HeII                           & 1640.42                                            &           &           &  \\
\textrm{O}~\textsc{iii} & 1660.81                                            &           &           &  \\
\SiIII                          & 1892.03                                            &           &           &  \\
\CIII                           & 1908.73                                            &           &           &  \\
\MgII                           & 2796.35,  2803.53                                      &           &           &  \\
\OII                            &  3727.09, 3729.88                                      &           &           &  \\
\NeIII                          & 3869.88, 3968.64                                      &           &           &  \\
\Heps                           & 3971.19                                            &           &           &  \\
\Hd                             & 4102.89                                            &           &           &  \\
\Hg                             & 4341.68                                            &           &           &  \\
\OIII                           & 4364.44                                            &           &           &  \\
\HeII                           &  4687.02                                            &           &           &  \\
\Hb                             & 4862.68                                            &           &           &  \\
\OIII                           & 4960.30, 5008.24                                      &           &           &  \\
\HeI                            & 5877.25                                            &           &           &  \\
\OI                             & 6365.54                                            &           &           &  \\
\Ha                             & 6564.61                                            &           &           &  \\
\SII                            & 6718.29, 6732.67                                      &           &           &  \\
\SIII                           & 9071.10, 9533.20                                      &           &           &  \\
\HeI                            & 10832.9                                           &           &           &  \\
\Pag                            & 10941.1                                           &           &           &  \\
\Pab                            & 12821.6                                           &           &           &  \\
\Paa                            & 18756.1                                           &           &           & \\
\cline{1-2}
\end{tabular}
\caption{Lines considered during the line-finding visual inspection effort. \label{tab:lines}}
\end{table}

\newpage
\section{Off-Center Emission}\label{sec:offcenter}
\begin{figure}[h]%\label{fig:redshiftspeccomp}
\includegraphics[width=1\columnwidth]{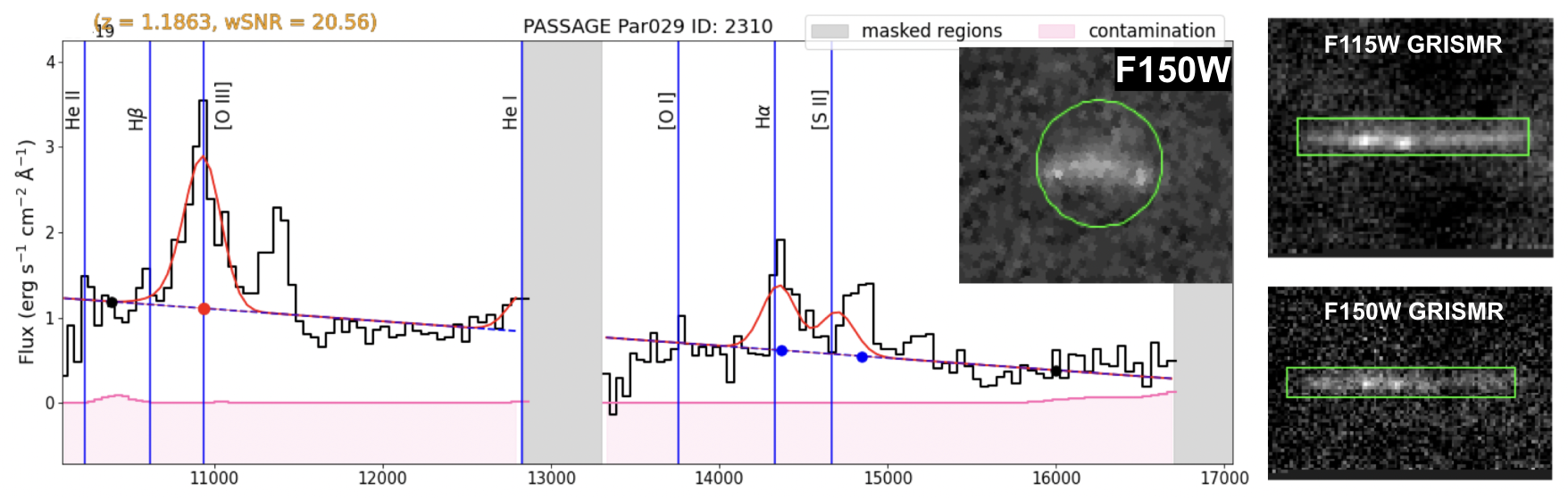}
\caption{Example of a PASSAGE source with off-center emission in Par029, which has F150W imaging (center insert) and F115W and F150W GRISMR spectra (right panels). The assigned spectroscopic redshift of the source varies depending on which set of peaks are assigned \Ha\ and \OIII, emphasizing the challenge of determining the redshift of sources with off-center emission. In the 1D spectra, the emission lines are aligned with the left peaks of \Ha\ and \OIII, which is likely offset from the true spectroscopic redshift. \label{fig:offcenter}}
\end{figure}

One additional challenge with the assignment of spectroscopic redshifts is off-center emission. Clumpy star-formation, galaxy mergers, accretion, and outflows may lead to emission peaks that do not correspond to the host galaxy's photometric center. An example of a source with substantial off-center emission is seen in Figure~\ref{fig:offcenter}. 
Two bright clumps can be seen in the F150W direct image, which translates to two distinct peaks for each emission line in both the 1D and 2D spectra. The two pairs of peaks likely correspond to \Ha\ and \OIII, but the spectroscopic redshift is likely inexact. If we assume the left peaks of each pair of emission line peaks are the observed wavelengths of \Ha\ and \OIII\ (as shown in figure~\ref{fig:offcenter}), $z_{spec}=1.1863$. If we instead assume the right peaks of each pair of emission line peaks are the observed wavelengths of \Ha\ and \OIII, $z_{spec}=1.2693$. It is most likely that neither of these redshifts best represent the true $z_{spec}$ of the source, and rather the true $z_{spec}$ lies somewhere between the two sets of emission lines. 
While these off-center emission scenarios are relatively rare, there will be increasingly large numbers of similar observations with upcoming Euclid and Roman spectroscopy. To disentangle the emission of off-center features and measure reliable redshifts, spectroscopic surveys with Euclid and Roman need to obtain observations where the traces are dispersed at multiple orientations, or roll angles.

\newpage
\newpage
\section{Expanded Comparison to Khostovan et al. 2025}\label{sec:khostcomp}
\begin{figure}[h]%\label{fig:redshiftspeccomp}
\includegraphics[width=1\columnwidth]{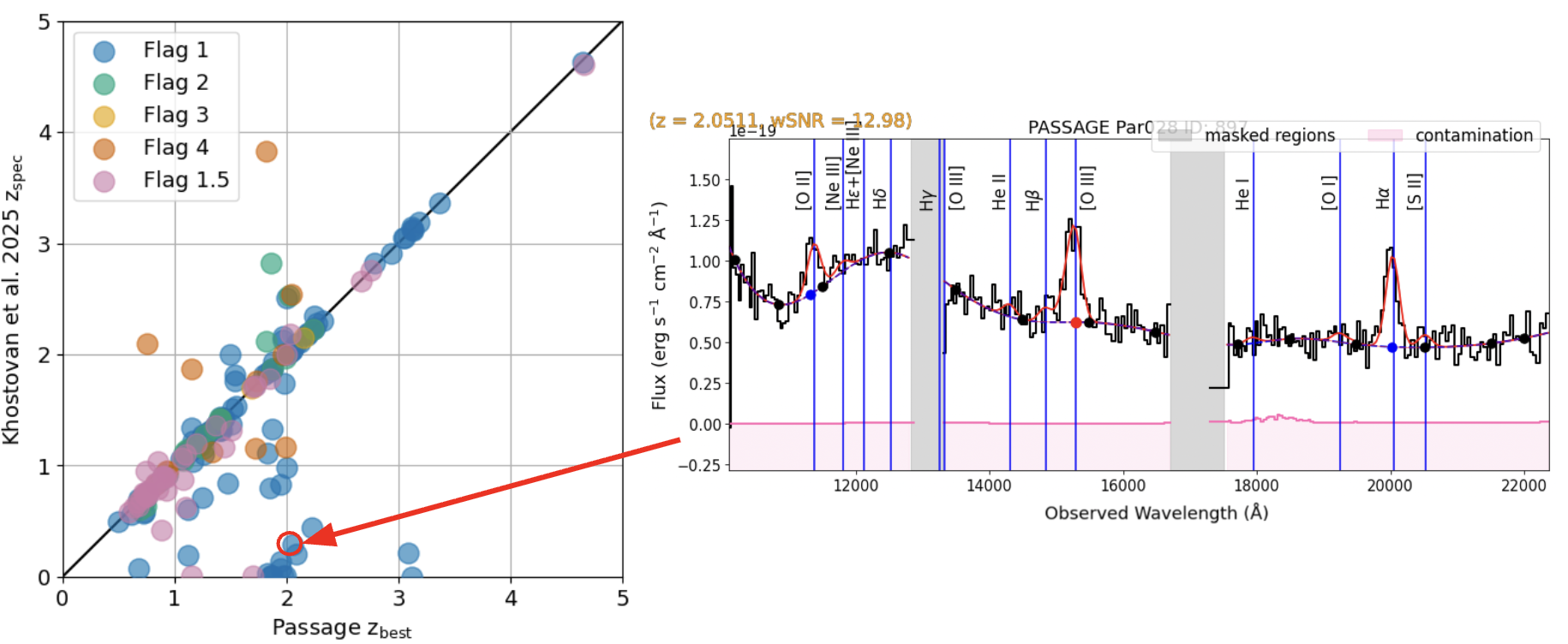}
\caption{Left panel: same as figure~\ref{fig:speczcomp}, except all spectroscopic redshifts from \citet{khostovan2025} and PASSAGE (regardless of quality flag) are included in this comparison. Right panel: an example source where PASSAGE can provide a more accurate spectroscopic redshift than was previously available. \label{fig:khostovancomp}}
\end{figure}

In section~\ref{sec:specredshiftcomp}, we conducted a redshift comparison between the PASSAGE spectroscopic redshifts and the \citet{khostovan2025} spectroscopic redshift compilation for redshifts with high quality flags in both samples. We expand this comparison to all 228 cross-matched sources between catalogs, regardless of quality flag. 
This comparison is seen in the left panel of figure~\ref{fig:khostovancomp}. 

168 of the 228 sources ($74\%$) agree within $|z_{Khostovan+}-z_{best}|<0.1$, indicating that even with the inclusion of sources with less secure redshift, the two catalogs are still in relatively strong agreement. This agreement drops from $74\%$ to $61\%$ if we exclude the high-quality redshift sources in both catalogs included in the comparison in Figure~\ref{fig:speczcomp}, indicating that even for low-quality redshifts, PASSAGE and \citet{khostovan2025} still agree a majority of the time.
Of the 60 sources with $|z_{Khostovan+}-z_{best}|>0.1$, 37 of them are PASSAGE flag 1 (robust) sources. For example, the source with $z_{best}=2.05113$ and $z_{Khostovan+}=0.28777$ (seen in the right panel of figure~\ref{fig:khostovancomp}) has a very robust PASSAGE redshift, constrained primarily by the \Ha, \OIII, and \OII\ lines. However, from the \citet{khostovan2025} compilation, it only has a `tentative' redshift measurement $\rm Q_f=1$ ($50\%$ confidence), meaning their spectroscopic redshift is very poorly constrained. The photometric $z_{COSMOS-Web}=1.95^{+0.08}_{-2.05}$ for this source agrees better with the PASSAGE $z_{best}$ than $z_{Khostovan+}$. This again emphasizes the importance of the PASSAGE sample: not only can PASSAGE substantially add to the existing sample of spectroscopic redshifts, but it can also increase the robustness of (and even correct) existing spectroscopic redshifts.

\end{appendix}

\newpage

\bibliographystyle{aasjournalv7}
\bibliography{PASSAGE_COSMOS}

\end{document}

%% file: obs_table.tbl
\begin{deluxetable*}{c|c|rrc|ccc|ccc|ccc}
\centering
\tablecaption{List of observed PASSAGE fields in Cosmos with their target coordinates and exposure times. The location column indicates whether the field is fully in Cosmos-Web (denoted `Cosmos-Web'), fully in COSMOS2020 (denoted `COSMOS2020'), or partially in Cosmos-Web and partially in the larger COSMOS2020 footprint ($>50\%$ indicates most of the field is in Cosmos-Web, $<50\%$ indicated that less than half of the field is in Cosmos-Web). SEP obj indicates the number of objects in each field that were identified in imaging by \sc SEP. CWT obj indicates the number of objects in each field that were identified as candidate emission-line sources by the CWT algorithm. Final obj indicates the number of objects in each field that were confirmed as line-emitters by the reviewers.}
\tablehead{
    \multirow{3}{*}{Field} &
    \multirow{3}{*}{Location} &
    \multirow{3}{*}{R.A. [deg]} &
    \multirow{3}{*}{Decl [deg]} &
    \multirow{3}{*}{Obs Date} &
    F115W & F150W & F200W & F115W & F150W & F200W &
    \multirow{3}{*}{SEP obj} & 
    \multirow{3}{*}{CWT obj} & 
    \multirow{3}{*}{Final obj}\\
    &&&&& Direct & Direct & Direct & Grism & Grism & Grism & \\
    &&&&& [hr] & [hr] & [hr] & [hr] & [hr] & [hr] & }
\startdata
    Par003 &  Cosmos-Web & 150.4103959 &      2.4051779 &     2022-12-30 &   -- &   -- & 0.62 &   --                 &   --                 & {\bf 1.24}\tnmark{a} & 3116  & 177 & 47\\
    Par005 &   Cosmos2020 & 150.6536980 &      2.0855800 &     2023-01-07 &   0.43 & 0.43 &   -- & {\bf 2.29}\tnmark{a}                 & {\bf 2.29}\tnmark{a}                 &   --                 & 4443  & 237 & 155 \\
    Par006 &   Cosmos2020 & 150.6762160 &      2.1293050 &     2023-01-06 &   0.43 & 0.43 &   -- & {\bf 2.29}\tnmark{a}                 & {\bf 2.29}\tnmark{a}                 &   --                 & 4323  & 276 & 169 \\
    Par017 &   Cosmos-Web & 150.3011088 &      1.8910817 &     2023-04-19 &   -- &   -- & 3.15 &   --                 &   --                 & 3.15                 & 4526  & 612 & 68 \\
    Par020 &   Cosmos2020 & 149.6229230 &      2.2251060 &     2023-04-23 &   -- & 0.13 &   -- & 4.87                 & 1.05                 &   --                 & 4186  & 283 & 176 \\
    Par023 &   Cosmos-Web & 150.0655286 &      2.5519213 &     2023-05-01 &   -- & 0.13 &   -- & 4.29                 & 1.05                 &   --                 & 4206  &260&150 \\
    Par024 &   Cosmos-Web & 150.1315237 &      2.4996595 &     2023-05-02 &   -- & 0.13 &   -- & 4.87                 & 1.05                 &   --                 & 3890 &208&119  \\
    Par025 &  Cosmos-Web &  150.2457730 &      1.9774739 &     2023-05-02 &   -- & 0.13 &   -- & 4.87                 & 1.05                 &   --                 & 3817  &275&143 \\
    Par026 &  Cosmos-Web &  150.1227170 &      2.4398773 &     2023-05-06 &   -- & 0.13 &   -- & 4.87                 & 1.05                 &   --                 & 3744 &231&152  \\
    Par028 &  Cosmos-Web &  150.0874070 &      2.4185307 &     2023-05-20 & 1.79 & 1.43 & 1.43 & {\bf21.26}\tnmark{a} & {\bf 7.08}\tnmark{a} & {\bf 5.37}\tnmark{a} & 3614 &894&285  \\
    Par029 &  Cosmos2020 &  150.2522870 &      1.7285570  &     2023-05-22 &   -- & 0.13 &   -- & 4.87                 & 1.05                 &   --                 & 3708  &218&123 \\
    Par049 &  Cosmos-Web ($<$50\%) &  149.7024380 &      2.0589470 &     2023-12-03 &   -- &   -- & 0.20 &   --                 &   --                 & 1.62                 & 2826&251&117\\
    Par051 &  Cosmos-Web ($>$50\%) &  150.5036633 &      2.2633940 &     2024-01-02 & 0.79 &   -- &   -- & {\bf 7.09}\tnmark{a} &   --                 &   --                 & 3473  &207&65 \\
    Par052 &  Cosmos-Web &  150.1552214 &      2.0418390 &     2024-01-03 &   -- & 2.04 & 2.04 & {\bf12.02}\tnmark{a} & 3.94                 & 3.94                 & 4433 &850&332  \\
    Par053 &  Cosmos-Web ($>$50\%) &  150.1030697 &      1.7931779 &     2024-01-04 &   -- &   -- & 0.20 &   --                 &   --                 & 1.62                 & 2971 &197 &83
    \\
\enddata
\tntext{a}{Total exposure is split over both R and C grism orientations. The remaining fields only have one orientation.}
\label{tab:obs_table}
\end{deluxetable*}